\documentclass[11pt,a4paper]{article}                               

\usepackage{latexsym}
\usepackage{mathrsfs}
\usepackage{amsthm}
\usepackage{url}
\usepackage{color}
\usepackage{jheppub}

\newcommand{\nc}{\newcommand}
\newcommand{\rnc}{\renewcommand}

\nc{\be}{\begin{equation}}
\nc{\ee}{\end{equation}}
\nc{\bea}{\begin{align}}
\nc{\eea}{\end{align}}

\rnc{\a}{\alpha}
\nc{\ab}{\bar{\a}}
\nc{\ap}{\a^{+}}
\nc{\abm}{\ab^{-}}
\rnc{\b}{\beta}
\nc{\bb}{\bar{\b}}
\nc{\bbp}{\bb_{\zb}^{+}}
\nc{\bm}{\b_{z}^{-}}
\nc{\oa}{\overline{\a}}
\nc{\ob}{\overline{\b}}
\rnc{\gg}{\gamma}
\rnc{\d}{\delta}
\nc{\f}{\phi}
\nc{\fb}{\bar{\phi}}
\nc{\vf}{\varphi}
\nc{\p}{\psi}

\rnc{\c}{\chi}
\nc{\la}{\lambda}
\nc{\m}{\mu}
\nc{\n}{\nu}
\rnc{\o}{\omega}
\nc{\Om}{\Omega}
\rnc{\t}{\theta}
\nc{\eps}{\epsilon}
\rnc{\S}{\Sigma}
\nc{\F}{\Phi}
\nc{\del}{\partial}

\nc{\trac}[2]{{\textstyle\frac{#1}{#2}}}

\nc{\ex}[1]{\mbox{e}^{\,\textstyle#1}}

\nc{\mat}[4]{\left(\begin{array}{cc}#1&#2\\#3&#4\end{array}\right)}

\nc{\som}[9]{\left(\begin{array}{ccc}#1&#2&#3\\#4&#5&#6\\#7&#8&#9%
\end{array}\right)}

\nc{\tr}{\mathop{\mathrm{tr}}\nolimits}
\nc{\ad}{\mathop{\mathrm{ad}}\nolimits}
\nc{\Tr}{\mathop{\mathrm{Tr}}\nolimits}
\nc{\Det}{\mathop{\mathrm{Det}}\nolimits}
\nc{\rk}{\mathop{\mathrm{rk}}\nolimits}
\nc{\ra}{\rightarrow}
\nc{\Ra}{\Rightarrow}
\nc{\LRa}{\Leftrightarrow}
\nc{\ot}{\otimes}
\nc{\nul}{\noindent\underline}
\nc{\non}{\nonumber\\}

\rnc{\lg}{\mathfrak{g}}
\nc{\lt}{\mathfrak{t}}
\nc{\lk}{\mathfrak{k}}
\nc{\lh}{\mathfrak{h}}
\nc{\Ad}{{\mbox{Ad}}}
\nc{\CC}{\mathbb{C}}
\nc{\EE}{\mathbb{E}}
\nc{\RR}{\mathbb{R}}
\nc{\ZZ}{\mathbb{Z}}
\nc{\NN}{\mathbb{N}}

\newtheorem{theorem}{Theorem}[section]

\newtheorem{example}[theorem]{Example}

\title{Massive Ray-Singer Torsion and Path Integrals}
\author[a]{Matthias Blau}
\author[b,c]{Mbambu Kakona}
\author[c]{George Thompson}

\affiliation[a]{Albert Einstein Center for Fundamental Physics\\ 
Institute for
Theoretical Physics, University of Bern, Switzerland}
\affiliation[b]{East African Institute for Fundamental Research (EAIFR)\\
University of Rwanda, Kigali, Rwanda}
\affiliation[c]{Abdus Salam International Centre for Theoretical Physics\\
Strada Costiera 11 Trieste, Italy}

\emailAdd{blau@itp.unibe.ch}
\emailAdd{kakona@eaifr.org}
\emailAdd{thompson@ictp.it}

\abstract{Zero modes are an essential part of topological field theories, but
they are frequently also an obstacle to the explicit evaluation of
the associated path integrals.  In order to address this issue in
the case of Ray-Singer Torsion, which appears in various topological
gauge theories, we introduce a massive  variant of the Ray-Singer
Torsion which involves determinants of the twisted Laplacian with
mass but without zero modes. This has the advantage of allowing one
to explicitly keep track of the zero mode dependence of the theory.
We establish a number of general properties of this massive Ray-Singer
Torsion. For product manifolds $M=N \times S^1$ and mapping tori
one is able to interpret the mass term as  a flat $\mathbb{R}_{+}$
connection and one can represent the massive Ray-Singer Torsion as
the path integral of a Schwarz type topological gauge theory. Using
path integral techniques, with a judicious choice of an algebraic
gauge fixing condition and a change of variables which leaves one
with a free action, we can evaluate the torsion in closed form. We
discuss a number of applications, including an explicit calculation
of the Ray-Singer Torsion on $S^1$ for $G=PSL(2,R)$ and a path
integral derivation of a generalisation of a formula of Fried for
the torsion of finite order mapping tori.}

\global\parskip=4pt

\begin{document}
\maketitle


\section{Introduction}

Many field theories suffer from issues related to zero modes which, in
particular, complicate the explicit evaluation of path integrals. This
is particularly
true in topological field theories where the zero mode structure is
all important. This obstacle appears even in the simplest topological
field theories such as $BF$ theories or Schwarz type theories in
general. In this paper we suggest a method for dealing with these issues by
introducing a massive version of Ray-Singer Torsion where all zero
modes are lifted.

$BF$ theories are a very simple class of topological gauge theories, 
and they have been with us for over 30
years \cite{BT-RST, Horowitz, WBF, BT-BF}. 
They derive their name from the fact that their classical action is
\begin{equation*}
S_{BF}= \int_{M} \Tr{B \wedge F_{A}} 
\end{equation*}
with $F_A$ the curvature 2-form of a connection $A$ and 
$B$ a Lie algebra valued $(\dim M - 2)$-form. Despite their 
(deceptive) simplicity, there have been few exact evaluations of the 
associated partition function on compact closed manifolds. 

One case which is completely understood is that of two-dimensional
$BF$ theory with a compact gauge group, for which the partition
function is the zero-coupling limit of that of two-dimensional Yang-Mills
theory, which can be evaluated in closed form in a variety of ways
\cite{Witten-2d1, Witten-2d2, Rusakov, BT-YM2,BT-Trieste-1993}. 

In dimension 3, $BF$-theories with gauge group $G$ can equivalently
be regarded as Chern-Simons gauge theories with the non-compact
gauge group $TG \simeq G \times \lg$, and in dimension greater than 3,
$BF$
theories have an intricate non-compact and moreover reducible
symmetry structure.  As a consequence there are different types of
zero modes and various sources of potential divergences.  Thus,
even though the construction and properties
of the quantum action for these theories
are well understood \cite{BT-BF,BBRT-PR,BT-Metric}, even in the 
case of 3-dimensional manifolds there is a dearth of examples of
exact evaluations.

Nevertheless, these theories are of interest from a number of points
of view. For example, from a mathematical perspective in the
3-dimensional case Witten \cite{WBF} has, using semi-classical
arguments, shown how they are related to the Johnson invariants
\cite{Johnson}, and it would certainly be desirable to develop tools
to explicitly evaluate the partition function in that situation. From a physics
perspective, $BF$ theories are e.g.\ related to formulations of theories
of gravity in various dimensions (see for example \cite{CGM,
Freidel-Speziale} for reviews), and again it would be of interest to gain a
better understanding of the corresponding partition functions in
those situations that goes beyond purely formal considerations. While we will not be dealing explicitly with 
$BF$-theories in this paper, the techniques that we develop here
(in a simpler context to be described next) are also applicable
to $BF$-theories. 

Some of the problems that beset $BF$-theories also arise in an 
even simpler class of theories, which have been with us even 
longer. There are the prototypical `Schwarz-type' 
\cite{Schwarz-Partition, Schwarz, Schwarz-Tyupkin} 
topological field theories 
with action 
\[
S_{RST} = \int_{M}\Tr{\mathscr{B}\wedge d_{A} \mathscr{C}} 
\]
Here $A$ is a flat connection on a vector bundle $\EE$, and 
the fields $(\mathscr{B},\mathscr{C})$ 
are $\EE$-valued differential $p-$ and $(\dim M -p-1)$-forms
respectively. Using covariant gauge fixing conditions for all the classical
and quantum fields that appear (i.e.\ $d_A*\mathscr{B}=d_A*\mathscr{C}=0$ etc.) 
with respect to some metric $g_M$ on $M$, these actions 
can be seen to provide a field-theoretic realisation of the Ray-Singer Torsion
\cite{Ray-Singer}  of $(A,\EE)$, given by 
\[
\tau_{M}(A,\mathbb{E},g_M)= 
\prod_{i=0}^{n}
\left(\Det_{\Omega^{i}(M,\mathbb{E})}{(\Delta_{A})}\right)^{(-1)^{i+1}i/2} 
\]
where $\Delta_{A}$ is the twisted Laplacian with respect to the metric $g_M$
acting on $\EE$-valued differential forms 
(for a more detailed description and definition of $\tau_M$ 
see the next section). 
These $\mathscr{B}-\mathscr{C}$ models are a kind of `Abelian'
version of $BF$ theories depending only on a fixed flat background
connection $A$, and they are Gaussian (quadratic, free) theories, 
so one might expect the evaluation of the corresponding partition
functions to be reasonably straightforward. 

Surprisingly enough, even in this case there are
essentially no intrinsically path integral determinations of the
partition functions.  The Ray-Singer Torsion can of course  be
calculated either by explicit calculations of the determinants
appearing in its definition (as in the classic calculation of Ray
for Lens spaces \cite{Ray}), or by actually calculating the
Reidemeister Torsion 
and appealing to the
equivalence of the Ray-Singer Torsion and the Reidemeister Torsion
conjectured by Ray and Singer \cite{Ray-Singer} and established in
\cite{Cheeger-1,Cheeger-2,Muller} (see e.g.\ \cite{Freed, Witten-2d1} for some 
sample calculations along these lines). 
Nevertheless, as mentioned, there are
essentially no intrinsically path integral derivations which make
use of the latitude afforded by that formalism, with the exception
of the Ray-Singer Torsion on $S^{1}$, which was explicitly calculated
in \cite{BT-CS} using purely path integral methods.

One of the factors contributing to this is the fact that, here too,
in general there are zero mode issues that need to be addressed.
Another one is the absence of a suitable alternative calculationally
efficient gauge condition that would allow one to derive an alternative
representation of the Ray-Singer Torsion from the path integral.
We will address these (and some related) issues in this paper. 

At a mathematical level, the zero mode issue is reflected in the
fact that in the presence of harmonic modes the correct definition
of the Ray-Singer Torsion (as a metric independent `topological'
quantity) requires a (metric dependent) cohomological correction
term on the cohomology, first described by Ray and Singer in
\cite{Ray-Singer-Analytic}.

At the field theory level, these harmonic modes correspond to zero
modes of the action and make the partition function ill-defined
(zero or infinity and multiple products thereof). In \cite{BT-BF} 
a BRST-invariant method to project out these harmonic
modes was proposed and 
outlined. As an aside (although an interesting aside, we believe), we
shall show here that the result of this prescription agrees precisely
with that obtained from the procedure advocated by Ray and Singer (Section \ref{subsubharmpi}).
Even so, while it is pleasing to see this agreement at the formal
level, this is still not a particularly useful prescription in realistic
cases, in particular for the purposes of explicitly evaluating the
partition function.

Therefore, in this paper we propose an alternative method of dealing
with the zero modes which turns out to be calculationally efficient
(and also extends to $BF$-theories). Namely, we pursue the seemingly
naive idea that, instead of projecting out the harmonic modes,  we
simply `lift' them by adding mass terms (Section \ref{submrst}).
This is of course not a totally original idea. In particular, it
is closely related to the  prescription of using a mass to regularise
the determinants that two of us advocated  on several occasions in
the past \cite{BT-CS,BT-S1,BT-Seifert} in order to lift certain
degeneracies in Chern-Simons theory (however, here we assign a more
fundamental role to this procedure and analyse its consequences).
On 3-manifolds of the form $\Sigma \times S^1$, it is also related
to the elegant equivariant prescription in complex Chern-Simons
theory of Gukov and Pei \cite{Gukov-Pei} which was motivated by
other geometric considerations.

In the case at hand, at the mathematical level 
this procedure simply amounts to replacing the Laplacians $\Delta_A$ that
enter the definition of the Ray-Singer Torsion by their `massive' versions
$\Delta_A + m^2$ (which have no zero modes). We refer to the object
that one obtains in this way, 
not very originally, as the Massive Ray-Singer Torsion. 
Up to a (judiciously chosen, see (\ref{mass-RST}))
constant prefactor, the Massive Ray-Singer Torsion is 
thus defined by
\[
\tau_{M}(A, \mathbb{E}, g_M,m)\propto \prod_{i=0}^{n} 
\left(\Det_{\Omega^{i}(M,\mathbb{E})}{(\Delta_{A}+ m^{2}
    )}\right)^{(-1)^{i+1}i/2} 
\]
If all the cohomology groups are trivial one can take the
$m^{2}\rightarrow 0$ limit and reproduce the usual Ray-Singer
Torsion. The interest thus lies in the case when the cohomology
groups are non-trivial. We will show that, perhaps somewhat 
unexpectedly, this massive Ray-Singer Torsion has a number of 
attractive properties in common with its `massless' counterpart
(including the triviality of the torsion in even dimensions, and 
a simple product formula). 

At the field theory level, there are a number of issues that need
to be addressed. First of all, in general one cannot expect to have
a massive deformation of the above 
$\mathscr{B}-\mathscr{C}$ action that
preserves the underlying gauge invariance (and hence gives rise to
the ratio of determinants entering the definition of the massive
torsion upon gauge fixing these symmetries, say). Secondly, even
if we had such an action, a priori we would not have a way to
evaluate the corresponding partition function on a general manifold
any more than we did with the original definition.  

However, as we will show, this situation brightens up considerably
for certain classes of manifolds, including for $S^{1}$, product
manifolds of the form $M= N \times S^1$ as well as, more generally,
mapping tori of $N$ which are fibrations $N\ra M\ra S^1$ over $S^1$
with fibre $N$. Also in the latter cases, a central role is played
by the Ray-Singer Torsion on $S^1$ itself, which we therefore look
at in detail (Section \ref{mrsts1}). 

In particular, we  will show that on $S^1$ the mass term can
be interpreted as the coupling to a flat $\mathbb{R}_{+}$ connection
$A^{\mathbb{R}}$ (Section \ref{mr+c}), and the massive Ray-Singer
Torsion for a flat gauge field $A$ can equivalently be regarded as
the standard (metric independent) Ray-Singer Torsion for a flat
gauge field $A + A^{\mathbb{R}}$ without zero modes. We determine
the Ray-Singer Torsion for $G \times \mathbb{R}_+$ connections 
(Section \ref{RSTM}), and show how it succinctly encodes the 
Ray-Singer Torsions for $G$ with different numbers of zero modes
(Section \ref{MDA}). As an example, we calculate the Ray-Singer Torsion
for $G=PSL(2,\mathbb{R})$ from its massive counterpart, and show
complete agreement with the calculation 
of the Reidemeister torsion by Stanford and Witten in \cite{Stanford-Witten}.
We also introduce a path integral representation of the Ray-Singer
Torsion on $S^1$, and evaluate it in closed form by a variant of
the method originally used in \cite{BT-CS}, namely by discretisation
and a suitable change of variables that trivialises the action.

Turning to higher-dimensional manifolds, we show that the gauge-theoretic
realisation and interpretation of the massive Ray-Singer Torsion
extends to manifolds of the form $M= N \times S^1$ (Section
\ref{mrstns1}) and to mapping tori (Section \ref{sec-map}).
In particular, we will show that the absence of zero modes allows
one to make a calculationally efficient gauge choice adapted to the
geometry of the situation (Section \ref{subagf}).  This choice of
gauge is a generalisation of the familiar `temporal' gauge $A_{0}=0$.
On a manifold of the type $N \rightarrow M \rightarrow S^{1}$
described above, the $p-$form field $\mathscr{B}$ has a component
$B^{p-1}d\theta$ along the $S^1$, and the generalised temporal gauge
is the algebraic gauge condition $B^{p-1}=0$.  With this gauge
choice, the path integral calculation of the partition function
simplifies significantly and boils down to knowing the torsion on
the circle $S^1$ (and that one is known and can be determined in a
variety of ways from first principles) and the application of various
local index theorems (Section \ref{PI-Massive-RST}). Both ingredients
have appeared previously in our calculation of Chern-Simons theory
partition functions on Seifert 3-manifolds \cite{BT-CS, BT-S1,
BT-Seifert, BKNT}. 

This will also allow us to give a path integral derivation of a
generalisation  of a formula due to Fried \cite{Fried} for the
torsion of a finite order mapping torus without needing to appeal
to the equivalence of Ray-Singer Torsion with Reidemeister torsion
\cite{Cheeger-1, Cheeger-2, Muller} (Section \ref{pidgff}).  We
should note here that a first attempt in this direction \cite{BJT-MT}
did not get very far calculationally, among other things
because it attempted to deal with
all, and not just finite order, diffeomorphisms.

In addition to its calculability (on mapping tori), there are a
number of other advantages to using the massive Ray-Singer Torsion.
Of particular importance is the ability to simultaneously keep track
of zero modes of differing orders and their corresponding torsions
(as seen concretely in the calculations of Section \ref{RSTM} and
\ref{MDA}).  Moreover, as we will see, the algebraic generalised
temporal gauge allows one to introduce the ghost system without the
need of picking a metric on $N$ at all. The topological nature of
the theory is then assured from the outset (Section \ref{subsubmiya}).

Yet another benefit of using the massive Ray-Singer Torsion
is that it provides an explanation of a rather puzzling fact that
arises within the Abelianisation / Diagonalisation programme
\cite{BT-CS,BT-Trieste-1993,BT-DIA}.  In particular,
in the evaluation of the Chern-Simons partition function on various
classes of 3-manifolds in \cite{BT-CS, BT-S1, BT-Seifert, BKNT},
one finds a localisation of the path integral onto what appear to
be non-flat connections (and Ray-Singer torsions associated to
them). Using the insight gained from the calculations that 
we have performed, we can now explain how to 
understand and interpret this (Section \ref{subsubabel}).

\section{Ray-Singer Torsion Revisited}

We begin this section by briefly reviewing the definitions of the Analytic Torsion and the
Ray-Singer Torsion \cite{Ray-Singer, Ray-Singer-Analytic} including the treatment
of harmonic modes in both cases. The standard Schwarz
type path integral representation of the Ray-Singer Torsion will also
be reviewed. We expand on our suggestion in
\cite{BT-BF} for dealing with non-trivial cohomology and establish
that the outcome of this procedure is equivalent to the prescription
of Ray and Singer.

This is followed by introducing `massive
Ray-Singer Torsion' where a mass is added to the
Laplacian appearing in the definition of the Ray-Singer Torsion. The
addition of the mass lifts the zero modes and provides one with a different
method of calculation. We establish a number of general properties of
the massive Ray-Singer Torsion which parallel those of the standard
torsion.

These ideas are applied in the case that the base manifold is the
circle. On $S^{1}$ both the Analytic Torsion as well as the Ray-Singer Torsion
with mass are determined. We show that the addition of a mass can be
understood equivalently as the introduction of a flat $\mathbb{R}_{+}$
connection. One may view the results either as a regularisation
procedure for a flat $G$ connection or as a bona fide evaluation of
the torsion for a flat $G\times \mathbb{R}_{+}$ connection (there are
no zero modes for this connection).

In \cite{BT-CS} we gave a path integral
derivation of the Ray-Singer
Torsion of a flat connection on $S^{1}$ in the adjoint representation, 
by projecting out the non-zero cohomology. Here we introduce the
$\mathbb{R}_{+}$ connection directly in the Schwarz type path integral. This
path integral is easily evaluated by a field redefinition that
exchanges the connections for twisted boundary conditions, leaving us with otherwise
an essentially free action.

\subsection{Analytic Torsion and Ray-Singer Torsion}

\subsubsection{Basic Definitions}

We quickly recall some basic definitions. Let $M$ be a compact oriented 
$n$-dimensional manifold equipped with a Riemannian metric $g_M$, and 
$\mathbb{E}$ a (real) vector bundle over $M$ with fibre $E$, and equipped with a flat 
connection $\nabla$ (corresponding to a gauge field $A$). We also 
assume that $\nabla$ is compatible with a fixed positive definite fibre 
metric  $\langle \cdot,\cdot\rangle$ on $\mathbb{E}$ (but we will keep this fibre metric 
fixed throughout the discussion and suppress it from the notation). 
These data define
\begin{itemize}
\item the nilpotent exterior covariant derivative $d_A$ on $\mathbb{E}$-valued 
differential forms, 
\be
d_A:\; \Omega^k(M,\mathbb{E})\ra \Omega^{k+1}(M,\mathbb{E}) 
\ee
with $(d_A)^2=0$, 
and the corresponding cohomology groups $H_A^k(M, \mathbb{E})$;
\item
the adjoint $\delta_A = d_A^*$ of $d_A$ with respect to 
the Hodge scalar product 
\be
\label{Hodge}
<a,b> = \int_M \langle a\wedge *b\rangle \;\;,
\ee
on $\Omega^k(M,\EE)$;
\item
the Laplace operators $\Delta_A = \Delta_A^{(k)}$ on $\EE$-valued $k$-forms, 
\be
\Delta_A = (d_A + \delta_A)^2 = d_A\delta_A + \delta_A d_A:\quad 
\Omega^k(M,\EE) \ra \Omega^k(M,\EE)
\ee
\end{itemize}
The spectrum of $\Delta_A$ is positive semi-definite, and by an analog of the usual Hodge
decomposition theorem one has an isomorphism between the harmonic modes 
(zero modes) of the Laplace operator and the cohomology groups, 
\be
\mathrm{Ker}\Delta_A^{(k)} \simeq H_A^k(M,\EE)\;\;.
\ee
In this setting, the \textit{Analytic Torsion} $\tau_M(A,\EE,g_M)$ is defined 
in terms of suitable ratios of determinants of these Laplace operators. 
In order to make sense of such infinite-dimensional determinants, 
Ray and Singer made use of a 
$\zeta$-function regularisation of these determinants \cite{Ray-Singer}.

When there are no zero modes (i.e.\ the cohomology groups are trivial - one also says that the complex defined by $d_A$ is acyclic), 
the determinant of the Laplace operator 
$\Delta_A$ can be defined via 
analytic continuation of the spectral $\zeta$-function
\be
\zeta_{A}(s) = \sum_{\mathrm{Spec}(\Delta_{A})} \lambda^{-s}
= \frac{1}{\Gamma(s)} \int_{0}^{\infty} t^{s-1}
\Tr{\left(\exp{\left(-t \Delta_{A}\right)}  \right)} dt
\ee
by setting 
\be
\Det(\Delta_A) = \ex{-\zeta_A^\prime(0)}\;\;.
\ee
We will also use the notation $\zeta_{A}^{(k)}(s)$ to denote the spectral 
$\zeta$-function of the operator $\Delta_A$ acting on the space 
$\Omega^k(M,\EE)$ of $k$-forms. 

When there are zero modes, 
Ray and Singer proposed
to use a $\zeta$-function regularisation that explicitly projects out the
harmonic modes, i.e.\
\be
\zeta_{A}(s)  =  \frac{1}{\Gamma(s)} \int_{0}^{\infty} t^{s-1}
\Tr{\left(\exp{\left(-t \Delta_{A}\right)}  - P \right)} dt
\ee
where $P$ is a projector onto the spaces of zero modes,
say $\lim_{t\rightarrow
  \infty} \exp{\left(-t \Delta_{A}\right)}$.

In either case, the \textit{Analytic Torsion} $\tau_M(A,\mathbb{E},g_M)$ is now
defined as
\be
\tau_{M}(A,\mathbb{E},g_M)= 
\prod_{i=0}^{n}
\left(\Det_{\Omega^{i}(M, \mathbb{E})}{(\Delta_{A})}\right)^{(-1)^{i+1}i/2} 
=
\prod_{i=1}^{n}
\left(\Det_{\Omega^{i}(M, \mathbb{E})}{(\Delta_{A})}\right)^{(-1)^{i+1}i/2} \label{RST-def}
\ee
For later convenience, we also write this definition in logarithmic form as
\be
\log\tau_M (A,\EE,g_M) = \sum_{k=0}^n (-1)^k (k/2)\zeta_A^{(k)\prime}(0) \;\;.
\label{logrst}
\ee

E.g.\ in the 1-dimensional case $n=1$ one has (by Hodge duality) 
\be
n=1:\quad \tau_M(A,\EE,g_M) = 
\left(\Det_{\Omega^{1}(M, \mathbb{E})}{(\Delta_{A})}\right)^{1/2} 
= \left(\Det_{\Omega^{0}(M, \mathbb{E})}{(\Delta_{A})}\right)^{1/2} \;\;.
\ee
A key property of the Analytic Torsion is that, in the acyclic case, it is 
independent of the metric $g_M$ on $M$ \cite{Ray-Singer}, 
\be
d_A \quad\text{acyclic}\quad\Ra\quad 
\tau_{M}(A,\mathbb{E},g_M)=\widehat{\tau}_M(A,\EE)\;\;.
\ee
In the non-acyclic case, the presence of harmonic modes
introduces a metric dependence, which can be cancelled 
\cite{Ray-Singer-Analytic} 
by introducing
a suitable metric-dependent volume factor $\rho_H(g)$ on the space of 
harmonic modes (identified with the cohomology groups). We will 
recall the definition of $\rho_H(g_M)$ below. One then has
\be
\widehat{\tau}_M(A,\EE) = \rho_H(g_M) \tau_M(A,\EE,g_M)\;\;.
\ee
In either case, the metric independent quantity $\widehat{\tau}_M$ is
also usually referred to as the \textit{Ray-Singer Torsion}.

{\bf Notation:} In case the vector bundle is trivial $\mathbb{E}=
      M \times E$ we may simplify the notation and write $\tau_M(A, E,
      g_{M})$ for $\tau_M(A,\EE, g_{M})$ and $\widehat{\tau}_M(A, E)$
      for $\widehat{\tau}_M(A,\EE) $. In order not to unduly burden the notation, we will occasionally suppress 
the vector bundle $\EE$ from the notation 
(when it is clear from the context, or simply from specifying 
$A$ itself), and we will frequently also not 
indicate explicitly the dependence on the metric $g_M$ 
(whenever this is not our main concern at that time).

\subsubsection{Treatment of Harmonic Modes}
\label{subsubharmonic}

Here we first explain how to define the cohomological correction
term $\rho_H(g_M)$ following the 
prescription given originally by Ray and Singer \cite{Ray-Singer-Analytic}.
We do this in some detail in order to simplify the comparison with the 
BRST-based path integral treatment of harmonic modes to be discussed in
Section \ref{subsubharmpi}. 
We then briefly indicate how to translate this into the modern perspective
viewing the Ray-Singer Torsion as an element of a suitable determinant 
line on the cohomology.

We assume that the cohomology groups $H^i_A(M,\EE)\equiv H^i_A$ are
not all trivial, and at the outset we choose some reference basis
$h^{(i)}_a$ of $H^i_A$, with $a = 1,\ldots, \dim H^i_A$. Given a metric
$g_M$ on $M$, any such class $h^{(i)}_a$ has a unique harmonic 
representative $f^{(i)}_a$, i.e.\
\be
\Delta_A f^{(i)}_a = 0\quad,\quad [f^{(i)}_a] = h^{(i)}_a\;\;.
\label{harmfh}
\ee
In the following it will be 
convenient to consider a smooth 1-parameter family of metrics $g_M(u)$ 
on $M$, and to parameterise and indicate the dependence of any quantity
on the metric $g_M(u)$ simply by its dependence on the parameter $u$. 
Thus for each $u$ we have the space of $u$-harmonic forms 
$\mathcal{H}^i_A(u)$, with the $f^{(i)}_a(u)\in \mathcal{H}^i_A(u)$
furnishing a basis of $\mathcal{H}^{i}_A(u)$. 

Moreover, for each $u$ we let 
$\gamma^{(i)}_a(u)$ be an orthonormal basis of $\mathcal{H}^i_A(u)$ 
with respect to the Hodge metric \eqref{Hodge}, i.e.\ we require
\be
<\gamma^{(i)}_a,\gamma^{(i)}_b>(u) = \int_{M} \langle\gamma^{(i)}_{a}(u)
\wedge *_u \gamma^{(i)}_{b}(u)\rangle = \d_{ab}\;\;.
\label{ongamma}
\ee
For each $u$ we now have two sets of basis vectors for the finite-dimensional 
vector space $\mathcal{H}^i_a(u)$, namely
the $\{f^{(i)}_a(u)\}$ and the $\{\gamma^{(i)}_a(u)\}$. There must therefore 
exist some invertible matrix $S^{(i)}_{ab}(u)$ such that 
\be
\gamma^{(i)}_a(u) = S^{(i)}_{ab}(u) f^{(i)}_b(u)\;\;.
\label{harmS}
\ee
Then $|\det S^{(i)}(u)| \neq 0$ is independent of the choice of orthonormal
basis $\gamma^{(i)}_a(u)$, and Ray and Singer show that 
\be
\begin{aligned}
\log\widehat{\tau}_M(A,\EE)  &\equiv
\sum_{k=0}^n (-1)^k \left( (k/2)\zeta_A^{(k)\prime}(0) 
+ \log |\det S^{(k)}|(u) \right) \\
&= \log\tau_M(A,\EE,u) + \sum_{k=0}^n (-1)^k \log |\det S^{(k)}|(u) 
\end{aligned}
\label{tauhat}
\ee
is indeed independent of $u$, and thus independent of the metric. 
Thus the multiplicative correction factor $\rho_H(u)$ is 
\be
\rho_H(u) = \prod_{k=0}^n |\det S^{(k)}(u)|^{(-1)^k}\;\;.
\label{rhoH}
\ee
As already mentioned, this factor does not depend on the choice of 
orthonormal basis of $\mathcal{H}^i_A(u)$. It does however depend
(weakly) on the initial ($u$-independent) 
choice of basis $h^{(i)}_a$ of $H^i_A$: under 
a change of basis $h^{(i)}\ra L^{(i)}h^{(i)}$ for some linear 
(and $u$-independent) 
transformation matrix $L^{(i)}$ one has $f^{(i)}\ra L^{(i)}f^{(i)}$ 
and therefore $S^{(i)} \ra S^{(i)}L^{(i)}$ and
\be
\rho_H(u) \ra  \rho_H(u) \prod_{k=0}^n |\det L^{(k)}|^{(-1)^k}\;\;.
\ee
It is now easy to see that one can redefine the Ray-Singer Torsion
in such a way that it is independent of this choice of basis, at
the expense of defining it to be an element of a suitable determinant
line (rather than as a real number). 

First of all 
recall that, for any vector space $V$ its determinant line $\Det V$ is 
defined by 
\be
\Det{V} = \stackrel{\dim{V}}{\wedge} V 
\ee
and that any basis 
$\{e_{i} \}$ of $V$ defines an element 
\be
e_{1} \wedge \dots \wedge e_{\dim{V}} \in \Det{V}\;\;.
\ee
Given a metric (or scalar product) on $V$, up to a sign (choice of
orientation) an orthonormal basis with respect to this metric defines
a preferred element of $\Det V$. 
Given some reference basis $v_i$, with $e_i = Sv_i$ for some 
matrix $S$, by the definition of the determinant one has
\be
e_{1} \wedge \dots \wedge e_{\dim{V}} = (\det S) 
v_{1} \wedge \dots \wedge v_{\dim{V}}.  
\ee
Thus, with respect to the fixed reference basis $v_i$, 
this element of $\Det V$ can be identified with the real 
number $\det S$. 

In particular, an orthonormal basis
$\gamma^{(i)}_a(u)$ of $\mathcal{H}^i_A$ defines an element of 
$\Det H^i_A$. In this way, we can lift $\rho_H(g_M)$ to a well-defined
element 
\be
\widehat{\rho}_H(g_M) \in \stackrel{\mathrm{even}}{\wedge} \Det{\mathrm{H}^{\bullet}_{A}}\,
\stackrel{\mathrm{odd}}{\wedge}
\left(\Det{\mathrm{H}^{\bullet}_{A}} \right)^{*} 
\ee
and define the Ray-Singer Torsion as 
\be
\widehat{\tau}_M(A,\EE) = \widehat{\rho}_H(g_M) \tau_M(A,\EE,g_M)\in 
\stackrel{\mathrm{even}}{\wedge} \Det{\mathrm{H}^{\bullet}_{A}}\,
\stackrel{\mathrm{odd}}{\wedge}
\left(\Det{\mathrm{H}^{\bullet}_{A}} \right)^{*} 
\ee
(see e.g.\ \cite{Bunke} Theorem 2.11 or (79) in \cite{Mnev}). 
This is independent of a choice of reference basis, but 
given some reference basis $h^{(i)}_a$, this reduces to the 
expressions for $\rho_H(g_M)$ and $\widehat{\tau}_M$ 
given in \eqref{rhoH} and \eqref{tauhat}.

While this may sound rather abstract, as an example, for the untwisted case
the correctly normalised element of $H^0(M,\mathbb{R})$ (the constant 
functions) would be not the function 1  but rather 
\be
 \frac{1}{\sqrt{\mathrm{Vol}(M, g_M)}} \in \Det{\mathrm{H}^{0}}\label{norm-H0}
\ee
with $\mathrm{Vol}(M, g_M)$ the volume of $M$ given by the chosen
metric.

\begin{example}{Consider the untwisted Laplacian $\Delta = 
-R^{-2} d^{2}/d\theta^{2}$ on
      the circle $S^1$ with metric $g_{S^1}=R^{2}d\theta \otimes d\theta$ 
(with $0 \leq \theta < 2\pi$ and $*1 = Rd\theta$). It 
has eigenmodes $\ex{in\theta}$ corresponding to the  eigenvalues
      $R^{-2} n^{2}$ with $n\in \mathbb{Z}$. In particular, there is a zero mode
(the constant function) for $n=0$. 
A standard $\zeta$-function calculation (excluding the zero mode
$n=0$) yields
\be
\zeta_{\Delta}(s) = \sum_{n\neq 0} \lambda_n^{-s} = 2 R^{2s} \sum_{n>0} n^{-2s} \equiv 2 R^{2s} \zeta(2s) \;\;.
\ee
Thus, using $\zeta(0)= -1/2, \zeta^\prime(0)= -(1/2) \log(2\pi)$,
\be
\zeta_\Delta^\prime(0) = 2 \log R^2 \zeta(0) + 4 \zeta^\prime(0) = -\log R^2 - 2 \log 2\pi = -\log (2\pi R)^2\;\;.
\ee
Therefore the $\zeta$-function regularised determinant of the Laplace operator
is
\be
\Det\Delta = \ex{-\zeta^\prime_\Delta(0)} = (2\pi R)^2 \;\;.
\ee
The Analytic Torsion is the square root
of the determinant of this operator (with the zero mode removed), and thus
\be
\tau_{S^1}(A=0 \, d\theta,g_{S^1}) \equiv\tau_{S^1}(R) = 2\pi R\;\;.
\ee
Notice that
this is metric dependent, as expected (because we have removed the zero mode).
To determine the correction factor $\rho_H(R)$ (thinking
of $R$ as playing the role of the parameter $u$ in this example), we note that
the harmonic 0- and 1-forms on the circle are (up to normalisation) just the
constant function 1 and the 1-form $d\theta$, with Hodge norm
\be
\int 1*1 = 2\pi R \quad,\quad \int d\theta * d\theta = 2\pi R^{-1}
\ee
Therefore, an orthonormal basis of the harmonic modes
is provided by $(2\pi R)^{-1/2}$ (in agreement with 
\eqref{norm-H0} above) and $(2\pi/R)^{-1/2}d\theta$. 
This provides precisely the overall factor $\sim R^{-1}$ required to cancel 
the metric dependence $\sim R$ of the Analytic Torsion, but 
there are metric-independent factors that depend on the choice of 
reference basis for $H^0$ and $H^1$:
\be
\begin{aligned}
\{h^{(0)}=[1], h^{(1)}=[d\theta]\} & \quad\Ra\quad \rho_H(R) = (2\pi
R)^{-1/2} (2\pi/R)^{+1/2} = R^{-1} \\ 
&\quad\Ra\quad \widehat{\tau}_{S^1} (A=0d\theta) = \rho_H(R)\tau_{S^1}(R) = 2\pi\\
\{h^{(0)}=[1], h^{(1)}=[d\theta/2\pi]\} & \quad\Ra\quad \rho_H(R) =
(2\pi R)^{-1/2} (2\pi R)^{-1/2} = (2\pi R)^{-1} \\ 
&\quad\Ra\quad \widehat{\tau}_{S^1} (A=0d\theta) = \rho_H(R)\tau_{S^1}(R) = 1
\end{aligned}
\ee
}
\end{example}
Of course, by construction both choices lead to (equally valid)
metric independent results.  To be specific, in the following,
whenever dealing with the Ray-Singer Torsion on $S^1$, we make the
second choice, i.e.\ we normalise the Ray-Singer Torsion such that
$\widehat{\tau}_{S^1} =1 $ for the trivial connection,
\be
S^1: \quad \rho_H(R) = (2\pi R)^{-1} \quad\Ra\quad \widehat{\tau}_{S^1}(A=0d\theta) = 1\;\;.
\label{s1norm}
\ee

\subsubsection{Actions for the Ray-Singer Torsion in any Dimension}
Here we first quickly review the standard field theory actions
for the Ray-Singer Torsion, in order to fix notation and as 
background for subsequent discussions. We then discuss in some 
detail the BRST-based path integral approach to dealing with 
harmonic modes suggested  in \cite{BT-BF}, and 
establish the precise equivalence with the mathematical approach
described in Section \ref{subsubharmonic}.
The traditional 
field theory approach to Ray-Singer Torsion \cite{Schwarz-Partition,
  Schwarz, Schwarz-Tyupkin, BT-BF}
uses first and second
order actions and is based on the fact that $\Delta_{A} = TT^{\dagger}$ with
$T=*d_{A} + d_{A}*$, in this case $T=\pm T^{\dagger}$, and both are
understood to act on direct sums of 
spaces of forms. The starting action in such a
path integral presentation is
\be
S_{M}(A)= \int_{M} \langle \mathscr{B},  d_{A}
\mathscr{C}\rangle \label{RST-Action}
\ee
for $A$ a flat connection on a bundle $\mathbb{E}$ while $\mathscr{B}$ and
$\mathscr{C}$ are sections of associated bundles. For real vector
bundles this is enough. If $\mathbb{E}$  is a complex vector bundle we
add the complex conjugate of (\ref{RST-Action}) to obtain a real
action. In the following we suppress, unless needed explicitly, the
complex conjugate fields, though one should be aware that they are there.

The symmetry is
\be
\delta \mathscr{B} = d_{A} \Sigma \quad,\quad
\delta \mathscr{C} = d_{A} \Lambda
\label{qsym}
\ee
for $\Lambda$ and $\Sigma$ forms of appropriate type. The $\delta_{A}$
component of $T$ comes from the requisite (covariant) gauge
fixing of (\ref{qsym}). 
Indeed, on any manifold $M$ one can gauge fix the shift symmetries
\eqref{qsym} of the action 
by imposing the usual covariant gauge conditions 
\be
d_{A}*\mathscr{B} =0, \quad d_{A}*\mathscr{C} =0
\label{cov-gf}
\ee
on the fields $\mathscr{B}$ and $\mathscr{C}$, and this leads to the 
operator $\Delta_A = TT^\dagger$. 
As we will briefly recall later on, 
the complete quantisation of such a system involves the gauge fixing of 
a hierarchy of
symmetries and gauge fixing and ghosts for ghosts, arising from 
the reducibility of the gauge 
parameters, as in $\Sigma \ra \Sigma +  d_A\Sigma^\prime$ etc. 
Including all the analogous covariant 
gauge fixing and measure or ghost terms also for the two 
towers of gauge symmetries of gauge symmetries, it was shown in detail
in \cite{Schwarz} in terms of resolvents and in 
\cite{BT-BF,BBRT-PR} in terms of BRST-symmetry and the Batalin-Vilkovisky
procedure \cite{BV1983} that in the end 
the partition function 
\be
Z_M[A] = \int D\Phi \, \exp{\left(i \int_{M} \langle \mathscr{B},  d_{A}
  \mathscr{C}\rangle + \mathrm{gauge\; fixing\; and\; ghost \;
    terms}\right)} \label{PF-RST}
\ee
(where $\Phi$ represents all the classical and quantum 
fields to be integrated over)
of these theories reduces precisely to the 
ratio of determinants of Laplace operators that defines the Ray-Singer Torsion;
more precisely one has
\be
Z_M[A] = \tau_M(A,\EE,g_M)^{(-1)^{p-1}}\;\;.
\label{ZRST}
\ee

The partition function also has a classical gauge invariance 
\be
A \rightarrow g^{-1}Ag + g^{-1}dg, \;\; \Phi \rightarrow \rho(g)
\Phi, \label{usual-gt}
\ee
acting on both the dynamical fields 
$\{\Phi\} = \{\mathscr{B}, \mathscr{C}, \ldots\}$ and the 
background gauge field $A$, 
provided that the gauge fixing conditions of the shift
gauge symmetry (\ref{qsym}) are chosen to respect this symmetry
(and the covariant gauge fixing conditions do). 
Here $\rho(g)$ is the transformation in the relevant representation
for the quantum field in question and under which the measure $D\Phi$
is invariant.

\subsubsection{Treatment of Harmonic Modes in the Path Integral}
\label{subsubharmpi}

When there are harmonic modes, the action has  more symmetries 
than those given in \eqref{qsym} above. Indeed, the symmetries are then
shifts by $d_A$-closed rather than just by $d_A$-exact forms, 
\be
\delta \mathscr{B} = \Gamma_{A},  \quad d_{A}\Gamma_{A}=0, \quad
\delta \mathscr{C} = \Psi_{A}, \quad d_{A}
\Psi_{A} =0
\label{qsym2}
\ee
Here each $\Gamma_A$ can be split non-uniquely into a $d_A$-exact piece
and a form representing a non-trivial cohomology class,
\be
\Gamma_A = d_A\phi + \beta_A \quad\text{with} \quad [\beta_A] \in H^p_A
\equiv H^p_A(M,\EE)
\;\;.
\label{qsym3}
\ee
As in Section \ref{subsubharmonic}, we consider a 1-parameter 
family of metrics $g_M(u)$. 
For each $u$, gauge fixing of the 
shift by the $d_A$-exact piece $d_A\phi$ 
of $\Gamma_A$ can, as usual, 
be accomplished e.g.\ by the covariant gauge fixing conditions \eqref{cov-gf}
with respect to the metric $g_M(u)$. 
As far as the shift symmetries by cohomologically non-trivial 
$\beta_A$ are concerned, in
line with the definition of the Analytic Torsion or Ray-Singer Torsion, for
each $u$ and for every field that appears in the classical (or
quantum) action we would like to gauge fix these by projecting out
the $u$-harmonic piece of the field. 
Moreover, we would like to do so in such a
way that the BRST-operator itself is metric independent so that
(formally at least) BRST-invariance of the path integral implies
metric independence.

We will now show how this can be accomplished by following the prescription
suggested in \cite{BT-BF}.
We will then show that this leads
precisely to the cohomological correction term $\rho_H(g_M)$ in the
Ray-Singer Torsion discussed in detail in Section \ref{subsubharmonic}, 
i.e.\ we will show that \eqref{ZRST} gets modified precisely to 
\be
\text{BRST gauge fixing of harmonic modes} \quad\Ra\quad 
Z_M[A] = \widehat{\tau}_M(A,\EE,g_M)^{(-1)^{p-1}}\;\;.
\label{ZWRST}
\ee
In the path integral
the symmetries \eqref{qsym2} 
go over to BRST symmetries together with appropriate transformations 
for a whole tower of ghost,
anti-ghost and multiplier fields. The antighosts and the 
multiplier fields are naturally paired but with opposite statistics so
that their measures are naturally given. This means that we do not
have to be explicit about zero modes of these as they can be
`canonically' excised in pairs (we will briefly come back to this below). 
The ghosts  and ghosts for ghosts associated with the hierarchy of
reducible symmetries 
\be
\d\mathscr{B} = d_A \Sigma^{(p-1)}\quad,\quad
\d\Sigma^{(p-1)} = d_A \Sigma^{(p-2)}\quad,\quad \ldots
\ee
will each have a symmetry of the type (\ref{qsym2}, \ref{qsym3}). 
Denoting $\mathscr{B}$ and its hierarchy of 
ghosts collectively by $\Omega^i$ (of form-degree $i$, and 
with $\Omega^{p}=\mathscr{B}$ and $i =0, \dots, p$), 
we write the corresponding BRST symmetry as
\be
\begin{aligned}
Q\Omega^{i} & =  d_{A}\Omega^{i-1} + c^{a}\beta^{(i)}_{a} \quad\quad &Qc^a =0\\
Q\overline{c}^{a} & =  \tau^{a} &Q \tau^a=0 
\end{aligned}
\ee
Here 
the $\beta^{(i)}_a$ with $a= 1, \dots ,\dim H^{i}_{A}$ are a fixed 
choice of forms  such that the
\be
h^{(i)}_a = [\beta^{(i)}_a] 
\ee
are a fixed basis of $H^i_A$.  $c^a,\overline{c}^a$ and $\tau^a$ are 
constant (zero-form) ghosts and multipliers respectively. The 
ghosts have ghost number $p-i+1$. 
Let us note the following:
\begin{itemize}
\item The choice of basis of $H^i_A$ is 
precisely the same ingredient as the one that enters the Ray-Singer prescription 
described in Section \ref{subsubharmonic}. 
\item Moreover, nothing depends 
on the actual choice of the $\beta^{(i)}_a$ in their cohomology class, 
since any change $\beta^{(i)}_a \ra \beta^{(i)}_a + d_A g^{(i-1)}_a$ can be 
absorbed into a redefinition of the $\Omega^{i-1}$. 
\item The BRST operator defined above is independent of the metric $g_M(u)$
(that will enter only in the gauge-fixing conditions). 
\item
There is a similar system for $\mathscr{C}$ which ranges
from $i=0, \dots, \dim{M}-p-1$ (or dually $i=p+1, \dots, \dim{M}$). 
\end{itemize}
Apart from the usual covariant gauge fixing terms, 
we add to the action terms of the form
\be
\begin{aligned}
Q\int_{M} \langle \Omega^{i}, *_u \overline{c}^{a} \gamma^{{(i)}}_{a}(u)\rangle
= \int_{M}& \left(  (-1)^{p-i} \langle \Omega^{i}, *_u \tau^{a}
  \gamma^{i}_{a}(u)\rangle\right.\\
& + \left. \langle d_A \Omega^{i-1}+ c^{b}\beta^{{(i)}}_{b},*_u \overline{c}^{a}
\gamma^{{(i)}}_{a}(u)\rangle \right) \label{harmonic-gf}
\end{aligned}
\ee
Here, precisely as in Section \ref{subsubharmonic}, the $\gamma^i_{a}(u)$
form an orthonormal basis of $u$-harmonic forms, i.e.\ \eqref{ongamma}
\be
\Delta_A(u)\gamma^{(i)}_a(u) = 0 \quad,\quad 
\int_{M} \langle\gamma^{(i)}_{a}(u)
\wedge *_u \gamma^{(i)}_{b}(u)\rangle = \d_{ab}\;\;.
\label{ongamma2}
\ee
Let us take a look at the two terms in turn:
\begin{enumerate}
\item
The effect of the first term is to precisely project out the $u$-harmonic 
terms of $\Omega^i$. Indeed, decomposing the field $\Omega^i$ as 
\be
\Omega^i = \Omega^i_a(u) \gamma^{(i)}_a(u) + \text{non-harmonic terms}\;\;,
\ee
where the $\Omega^i_a(u)$ are constant on $M$, 
one sees that (due to the orthogonality of harmonic and non-harmonic terms in the scalar product) the latter do not contribute, and the first term is simply
\be
\int_M \langle \Omega^{i}, *_u \tau^{a} \gamma^{{(i)}}_{a}(u)\rangle 
= \Omega^i_b(u)\tau^a \int_M \langle\gamma^{(i)}_b(u), *_u \gamma^{{(i)}}_{a}(u)\rangle 
= \Omega^i_a(u)\tau^a\;\;.
\ee
Thus the integral over the multipliers $\tau_a$ will simply impose the 
requirement that the $\Omega^i_a(u)=0$,  i.e.\ that 
the $u$-harmonic part of $\Omega^i$ is equal to 0. It thus has 
the same effect as the projector $P$ has in the definition 
of the $\zeta$-function regularised determinants appearing in the 
Analytic Torsion. 

However, this 
part all by itself would not lead to a metric independent answer, 
just like projecting out the harmonic modes in the definition of the 
Analytic Torsion all by itself does not accomplish this. As we will
see, metric independence will be restored by including the second
(ghost) term, as would have been expected on the basis of BRST-invariance.

\item In the second (ghost) term, 
the $d_A$-exact piece $d_A\Omega^{i-1}$ 
does not contribute, because it is orthogonal to the
harmonic term it is paired with. In order to evaluate the remaining 
term
$\int_M \langle c^{b}\beta^{{(i)}}_{b},*_u \overline{c}^{a}
\gamma^{{(i)}}_{a}(u)\rangle$, 
we split $\beta^{(i)}_a$ as 
\be
\beta^{(i)}_a = f^{(i)}_a(u) + d_A g^{(i-1)}_{a}(u) 
\ee
where $f^{(i)}_a(u)$ is $u$-harmonic. This split is unique, because by Hodge 
theory every
cohomology class $h^i_a = [\beta^{(i)}_a]$ contains a unique $u$-harmonic 
representative. Note that this $f^{(i)}_a(u)$ is exactly the same as the one
that appeared in \eqref{harmfh}. 
By the same reasoning as just above, the $d_A$-exact piece
does not contribute, and thus one is left with the task of evaluating
$c^{b}\overline{c}^a \int_M 
\langle f^{{(i)}}_{b}(u),*_u \overline{c}^{a} \gamma^{{(i)}}_{a}(u)\rangle $.
By \eqref{harmS}, the bases $\{ 
f^{{(i)}}_{a}(u)\}$ and $\{\gamma^{{(i)}}_{a}(u)\}$ of $u$-harmonic forms
are related by the matrix $S^{(i)}(u)$, and therefore 
\be
c^{b}\overline{c}^a \int_M 
\langle f^{{(i)}}_{b},*_u \overline{c}^{a} \gamma^{{(i)}}_{a}(u)\rangle 
= (S^{(i)})^{-1}_{ab}c^b \overline{c}^a\;\;.
\ee
In this term, the ghost $c$ has ghost number $p-i+1$, and therefore from 
the ghost integral one obtains a contribution proportional to
\be
(S^{(i)})^{-1}_{ab}c^b \overline{c}^a \quad\Ra\quad 
|\det S^{(i)}|^{(-1)^{p-i+1}}\;\;.
\ee
\end{enumerate}
Collecting all the ghost terms, including those coming from the field
$\mathscr{C}$, one thus finds that the ghost determinants contribute
\be
\begin{aligned}
\log Z_M &= (-1)^{p-1}\log\tau_M + \sum_{i=0}^n (-1)^{p-i+1} \log|\det S^{(i)}|
\\
&=
(-1)^{p-1}\left( \log\tau_M + \sum_{i=0}^n (-1)^{i} \log|\det S^{(i)}|\right)
\\
&= (-1)^{p-1} \log \widehat{\tau}_{M}
\end{aligned}
\ee
as claimed, because the 2nd term in the 2nd line is precisely equal 
to the Ray-Singer correction term in \eqref{tauhat}. 

It remains to explain what to do with all the other (anti-ghost and
multiplier) fields of various form degrees that appear upon gauge
fixing the theory, but these are also easily dealt with \cite{BT-BF}: 
they always
come in pairs of $k$-forms 
$(\bar{\sigma}, \pi)$ with $Q\bar{\sigma}=\pi, Q\pi=0$. 
In order to gauge fix their harmonic modes to zero, we again introduce
a pair $(\bar{c}^a, \tau^a)$ of zero-forms, 
and simply add to the action the term
\be
Q\int_M \langle \bar{\sigma} *_u \bar{c}^a \gamma^{(k)}_a\rangle
= 
\int_M \langle \pi  *_u \bar{c}^a \gamma^{(k)}_a\rangle
\pm 
\int_M \langle \bar{\sigma} *_u \tau^a \gamma^{(k)}_a\rangle
\ee
which simultaneously projects out the $u$-harmonic modes of $\pi$ and 
$\bar{\sigma}$ in a manifestly BRST-invariant manner. 

Thus the path integral is now well-defined, and we have reproduced
on the nose the cohomological (or determinant line) correction
factor to the Ray-Singer Torsion, required to restore metric
independence in the presence of harmonic modes, from a BRST-invariant
gauge fixing of the harmonic modes in the path integral.

The prescription we have just described is, therefore, perfect for all
formal purposes; unfortunately, for practical calculations it is
somewhat cumbersome, as one would need to know the harmonic modes and a
more convenient way of handling them. This is one of the main reasons
for introducing the Massive Ray-Singer Torsion as it allows us to
by-pass that need.

\subsection{Massive Ray-Singer Torsion: Definition and General Properties}
\label{submrst}

Instead of projecting out the zero modes right away, we are able to follow them
around (and project them out later if so required). To lift the zero modes we 
add a small mass term to the
Laplacian, 
\be
\Delta_A \ra \Delta_A + m^2\;\;.
\ee
Since the spectrum of the twisted Laplacian $\Delta_A$ 
is positive semi-definite, 
adding the small mass lifts the zero-modes so that the spectrum of the
massive Laplacian, $\Delta_{A} + m^{2}$ is positive
definite.  Its spectral zeta function
\be
\zeta_{A}(s; m) = \frac{1}{\Gamma(s)} \int_{0}^{\infty} t^{s-1}
\Tr{\exp{\left(-t (\Delta_{A} + m^{2})\right)}} dt \label{mass-zeta}
\ee
then defines the determinant
\be
\Det{\left(\Delta_{A}+ m^{2}\right)} =\exp{\left(- \zeta_{A}'(0;m)\right)}
\;\;.
\ee
Formally one has
\be
\Det{\left(\Delta_{A}+ m^{2}\right)} = |m|^{2\dim{\mathrm{H}_{A}}}
\prod_{\lambda \in \mathrm{Spec}'} \lambda \, .\, (1 + \frac{m^{2}}{\lambda})
\ee
where $\mathrm{Spec}'$ is the non-zero spectrum of
$\Delta_{A}$. Rather more explicitly from the definition
(\ref{mass-zeta}) we have
\be
\zeta_{A}(s; m) = \zeta_{m}(s) + \widehat{\zeta}_{A}(s; m)
\ee
with
\be
\zeta_{m}(s)= \frac{1}{\Gamma(s)} \int_{0}^{\infty} t^{s-1}
\Tr{\left( P\; \exp{(-t m^{2})}\right)} dt \label{zeta-m}
\ee
and
\be
\widehat{\zeta}_{A}(s; m)= \frac{1}{\Gamma(s)} \int_{0}^{\infty} t^{s-1} 
\Tr{(1-P)\exp{\left(-t (\Delta_{A} + m^{2})\right)}} dt
\ee
The projection onto the non-zero modes of $\Delta_A$ can then 
be accomplished by considering 
\be
\begin{aligned}
  \lim_{m\rightarrow 0}\, \Det{\left.\left(\Delta_{A}+ m^{2}
      \right)\right|_{\Omega^{\bullet}_{A}\ominus
      \mathrm{H}^{\bullet}_{A}}}& \equiv  \lim_{m\rightarrow 0}\,
  \frac{\Det{\left.\left(\Delta_{A}+ m^{2} 
      \right)\right|_{\Omega^{\bullet}_{A}}}}{\Det{\left.\left( m^{2} 
      \right)\right|_{
      \mathrm{H}^{\bullet}_{A}}}}  \\
& =  \lim_{m\rightarrow 0}\,
  |m|^{-2\dim{\mathrm{H}^{\bullet}_{A}}} \Det{\left.\left(\Delta_{A}+ m^{2}
      \right)\right|_{\Omega^{\bullet}_{A}}} 
\label{Det-project}
\end{aligned}
\ee
Here $\Det{\left.\left( m^{2}  \right)\right|_{
    \mathrm{H}^{\bullet}_{A}}}$ is the determinant that arises on
using the $\zeta$-function $\zeta_m(s)$ (\ref{zeta-m}).

From this point of view the
mass regularised or massive Ray-Singer (or Analytic) Torsion is defined to be
\be
\tau_{M}(A, \mathbb{E}, g_M,m)= |m|^{e(M).\rk(\mathbb{E}). \dim{M}/2} \, \prod_{i=0}^{n} 
\left(\Det_{\Omega^{i}(M, \mathbb{E})}{(\Delta_{A}+ m^{2}
    )}\right)^{(-1)^{i+1}i/2} \label{mass-RST}
\ee
where $e(M)$ is the Euler characteristic of $M$. 
The ratio of
determinants is exactly the same as that in the definition of the
Analytic Torsion in \eqref{RST-def}. Thus it is only the prefactor
(which is only non-trivial in even dimensions) which may require
some justification. As we will explain below, this factor is chosen
such that the massive regularised Ray-Singer Torsion, just as the
standard Analytic Torsion itself, is trivial (equal to 1) for even
dimensional manifolds.

While the original Ray-Singer Torsion
$\tau_{M}(A, \mathbb{E},g_M).\rho_{H}(g_M)$ is independent of the choice of
Riemannian metric, in general 
$\tau_M(A, \EE, g_M,m)$ will not be (even though there are no zero modes of
the operator $\Delta_A + m^2$), and one should also not expect there to be 
a simple correction factor analogous to the cohomological correction term
$\rho_{H}(g_M)$ which would render the massive torsion 
metric independent in general. We will discuss in much more detail in 
subsequent sections, under which circumstances
the massive Ray-Singer Torsion can be 
understood or interpreted as defining a metric-independent quantity. For the 
time being, however, we note that  in general 
the massive torsion has a number of useful and interesting
properties
(some of which are identical to those of the standard (massless)
Ray-Singer or Analytic Torsion):

\begin{itemize}
\item 
By removing the prefactor and 
projecting out the harmonic modes of $\Delta_A$, as in 
\eqref{Det-project}, 
the standard Analytic Torsion can be recovered from the massive
torsion as 
\be
\tau_{M}(A,g_M) = \lim_{m\rightarrow 0}\,
|m|^{(\sum_{i=1}^{n}(-1)^{i} i\dim{\mathrm{H}^{i}_{A}} -
  e(M).\rk(\mathbb{E}). \dim{M}/2 )}\, . \tau_{M}(A,
g_M, m) 
\label{RST-limit}
\ee
\item
The alternating sum of weighted dimensions of the cohomology
groups that appears here is given by the derivative of the 
Poincar\'{e} polynomial 
\be
P_{A}(t) = \sum_{i=1}^{n} t^{i} \dim{\mathrm{H}^{i}_{A}}
\ee
at $t=-1$, 
\be
\left.\frac{dP_{A}(t)}{dt}\right|_{t=-1} = - \sum_{i=1}^{n}(-1)^{i}
i\dim{\mathrm{H}^{i}_{A}} \;\;.
\ee
Since the limit $m\ra 0$ in \eqref{RST-limit} is well-defined, 
one has an expansion of the massive torsion 
in non-negative powers of $m$ as
\be
\tau_{M}(A, \mathbb{E},g_M, m) = |m|^{(e(M).\rk(\mathbb{E}). \dim{M}/2+\dot{P}_{A}(-1))} \,
\sum_{n=0}^{\infty} a_{n}(M,A) m^{n} \label{reg-RST}
\ee
In particular, for $M$ even-dimensional one has 
\be
\dim M = 2k \quad\Ra\quad \dot{P}_A(-1) = - e(M).\rk(\EE)\dim M/2 
\ee
and therefore 
\be
\dim M = 2k \quad\Ra\quad 
\tau_{M}(A, \mathbb{E},g_M, m) = 
\sum_{n=0}^{\infty} a_{n}(M,A) m^{n} \;\;.
\label{reg-RST2}
\ee
As an aside we note that the quantity $\dot{P}_A(-1)$ naturally 
appears in a variety of contexts, and in some sense is an odd-dimensional 
analogue of the Euler characteristic - see for example \cite{BDN}.
\item 
If $M$ is even dimensional the massive Ray-Singer Torsion 
is actually trivial, that is it satisfies
      \be
      \tau_{M}(A, \mathbb{E},g_M, m) = 1 \label{even-RST}
      \ee
for all $m$. In particular  the coefficients
      in (\ref{reg-RST2}) are $a_{0}(M,A) = 1$ and $a_{i}(M,A)=0, \; i\geq 1$.

This is just as for the standard (massless) Analytic Torsion or Ray-Singer Torsion, 
which is trivial (and =1) in even dimensions in the acyclic case.
The proof relies on the fact that both $d_A$ and the Hodge operator 
$*$ commute with the
Laplacian $\Delta_A$, leading to a cancellation of non-zero eigenvalues
between $k$-forms and $(n-k-1)$-forms in even dimensions in the
expression for the Analytic Torsion. This argument extends verbatim
to the massive case, as also the operator $\Delta_A + m^2$ commutes
with both $d_A$ and $*$. An alternative argument, using a result of
\cite{BT-BF}, is given in Appendix \ref{mRST-props}.

    \item There is another property of the Ray-Singer Torsion that 
is also shared by its massive counterpart, namely a product formula.  
Suppose $M$ and $N$ are compact, closed and
      oriented manifolds and equip $M \times N$ with the product
      metric. Let $E$ be a vector bundle over $M \times N$ which
      allows for flat connections with local form $A \in \Omega^{1}(M, \lg)
      \otimes \Omega^{0}(N)$, up to gauge equivalence. Then
      \be
      \tau_{M\times N}(A, \mathbb{E},g_{M\times N}, m) = \tau_{M}(A, 
\mathbb{E},g_M, m)^{e(N)} \, . \, \tau_{N}(g_N,m)^{e(M)\rk{\EE}} 
\label{product-RST}
      \ee
The proof (Appendix \ref{mRST-props}) 
follows the original proof of Theorem 2.5 in Ray and Singer \cite{Ray-Singer} 
(with a slight generalisation we gave in \cite{BT-CS}).

In view of the triviality of the massive torsion for even-dimensional 
manifolds, and that of the Euler characteristic for odd-dimensional manifolds, 
this product formula is only of interest if $\dim M$ is odd and 
$\dim N$ is even. In this case the product formula (\ref{product-RST}) 
reduces to
\be
\tau_{M\times N}(A, \mathbb{E},g_{M\times N}, m) = \tau_{M}(A, 
\mathbb{E},g_M, m)^{e(N)} 
\label{product-RST2}
\ee
\item
Lastly, there are a number of attractive results, due to Fried \cite{Fried},
for the Reidemeister Torsion (and hence the Ray-Singer Torsion). The
one of interest to us here is the following:

Let $M$ be an odd dimensional mapping torus $N \rightarrow M
  \stackrel{\pi}{\rightarrow} S^{1}$
  with $f$ a finite order diffeomorphism acting on $N$. For $E$ a
  bundle over the base $S^{1}$ with connection $\phi d\theta$ where
  $\phi$ is constant and
  $f_{i}^{*}: \mathrm{H}^{i}(N,\mathbb{R}) \rightarrow \mathrm{H}^{i}(N,\mathbb{R})$. Then
  \be
\tau_{M}(\phi, \pi^{*}\mathbb{E}) =
\prod_{i=0}^{\dim N}\left|\det{(1-\rho(\ex{-2\pi\phi})\otimes
    f_{i}^{*})}\right|^{(-1)^{i}} \label{fried-map}
\ee
We will establish a somewhat stronger result (including the mass term)
in Section \ref{sec-map} using field theory techniques. 
    \end{itemize}

\begin{example}\label{mass-s1}{For the untwisted Laplacian with mass $m>0$ on $S^{1}$ one has
      \be
      \tau_{S^{1}}(R,m) = \frac{1}{m}\prod_{n\geq 0}R^{-2}(n^{2}+R^{2}m^{2})= \frac{1}{m}
      \prod_{n\geq 0}R^{-1}(n+ iRm)\prod_{n\geq 0}R^{-1}(n-iRm)
      \ee
which suggests making use of the Hurwitz $\zeta$-function $\zeta(s;a)
= \sum_{n\geq 0} (n+a)^{-s}$ ($a \in \mathbb{C}$ but not 0 or a
negative integer). We therefore define
\be
\tau_{S^{1}}(R,m) = \lim_{s\rightarrow 0} \frac{1}{m}.  \exp{\left(-d/ds.
    \left[R^{s}\zeta(s, iRm) + R^{s}\zeta(s, -iRm)
    \right]\right)}
\ee
Note that $\zeta(0, a) = 1/2-a$ and $\zeta'(0, a)
= \ln{\Gamma(a)}- \ln{(2\pi)}/2$. Hence
\be
\tau_{S^{1}}(R,m) = \frac{2\pi}{mR \, \Gamma(imR)\Gamma(-imR)} = 2i
\sin{(\pi (1+imR))} = 2\sinh{(\pi mR)}\;\;,
\ee
the second last equality arising from Euler's reflection formula. As $m
\rightarrow 0$ this vanishes, reflecting the presence of the zero mode.
Indeed we have
\be
\lim_{m \rightarrow 0} \tau_{S^{1}}(R,m) = 2\pi R m + \dots
\ee
where the exact same metric dependence is maintained, at this order, as in the
massless case.
}
\end{example}

\subsection{Massive Ray-Singer Torsion on $S^{1}$}\label{mrsts1}

Note that the massive spectrum determined in Example \ref{mass-s1}, 
with $n \ra n \pm iRm$, looks 
like it could also be produced by the coupling to a gauge field
$\sim iRm$, in which case the issue of metric (in-) dependence presents
itself in a slightly different way. 
We shall now reconsider the above example from this point of view and
then generalise it in a variety of ways.

\subsubsection{Mass and $\mathbb{R}_{+}$ Connections}\label{mr+c}

Thus consider once again the circle  
$S^{1}$ with local coordinate $0 \leq \theta < 2\pi $ and with 
a metric $g_{S^1}=R^2 d\theta \otimes d\theta$ such that 
$\mathrm{Vol}(S^{1}, g_{S^1}) = 2\pi R$, 
$*1= R d\theta$ and $*d\theta = R^{-1}$. However, rather than
looking at the untwisted Laplacian (as in the above Example),
consider now a vector bundle $\mathbb{E}$ over $S^{1}$, 
equipped with a (necessarily flat) connection $A=\phi d\theta$, 
with associated exterior covariant derivative
\be
d_{A} = d\theta \, D_{\phi}, \;\;\; D_{\phi} = \partial_{\theta} +
\phi
\ee
Take $\phi$ to be anti-Hermitian (with respect to the fibre metric $\langle \cdot,\cdot\rangle$). 
The Ray-Singer Torsion on $S^{1}$ of the flat connection
$A=\phi\, d\theta$ is formally the positive root
\be
\tau_{S^{1}}(\phi, \mathbb{E})= \left(\Det_{\Omega^{1}(S^{1},\EE)}{\Delta_{A}}\right)^{1/2} 
\ee
of the twisted Laplacian $\Delta_A$ on $\Omega^k(S^1,\EE)$, i.e.
$\Delta_{A}= -(*d_{A})^{2}$ on $\Omega^{0}(S^{1},\EE)$
and $\Delta_A= -(d_{A}*)^{2}$ on $\Omega^{1}(S^{1},\EE)$.
The spectra of the two Laplacians are the same (by Hodge duality, 
and $\Delta_A * = *\Delta_A$). The positive square root can then be
taken to be either $|\Det{(id_{A}*)}|$ or $|\Det{(i*d_{A})}|$ on
the appropriate space of forms,
\be
\tau_{S^{1}}(\phi, \mathbb{E})= 
\left|\Det_{\Omega^{1}(S^{1},\EE)}{(id_A*)}\right|=
\left|\Det_{\Omega^{0}(S^{1},\EE)}{(i* d_A)}\right|
\ee
Adding a mass term to the twisted Laplacian amounts to the substitution
\be
\Delta_{A} \rightarrow  \Delta_{A} + m^{2} = R^{-2}
(iD_{\phi}+imR)(iD_{\phi}-imR)\;\;.
\ee
Note that, since $i\phi \pm im R$ is the sum of a Hermitian and an anti-Hermitian
part, the individual factors on the right-hand side
 are neither Hermitian nor anti-Hermitian. 

However, at this point we can regard the mass term $mR$ as an $\mathbb{R}_{+}$
connection (or as the imaginary component of a $U(1)_{\mathbb{C}}$
connection). This means we take $\mathbb{E}\otimes \mathbb{L}^{\pm 1}$ to be a $G \times
\mathbb{R}_{+}$ bundle over $S^{1}$ with a flat connection 
$A^\pm = (\phi \pm \mu) d\theta$. The relationship between the connection
$\mu$ and the mass term $m$ is 
\be
\mu = R m \quad\LRa\quad \mu d\theta = * m\;\;.
\label{muRm}
\ee
In this case
the massive Ray-Singer Torsion takes the more symmetric form
\be
\tau_{S^{1}}(\phi, \mathbb{E}, m) = \sqrt{\tau_{S^{1}}(\phi+  \mu,
  \mathbb{E}\otimes \mathbb{L})\, \tau_{S^{1}}(\phi- \mu,
  \mathbb{E}\otimes \mathbb{L}^{-1})} \label{RST-fact}
\ee
where we have set
\be
\label{taupm}
\tau_{S^{1}}(\phi \pm  \mu,
\mathbb{E}\otimes \mathbb{L}^{\pm}) = \Det_{\Omega^{1}(S^{1}, \mathbb{E})}{\left(R^{-1}
    (iD_{\phi} \pm i \mu) \right)} 
\ee
The factorised form for the Ray-Singer Torsion (\ref{RST-fact}) agrees
with what was found in
\cite{BT-CCS} for complex Chern-Simons theory. In the following we fix
on one sign and consider for concreteness 
$\Det_{\Omega^{1}(S^{1},\mathbb{E}\otimes \mathbb{L})}{\left(R^{-1}(iD_{\phi} + i
    \mu)\right)}$, the other case is easily obtained.

\subsubsection{Metric Independence of the $U(1)_{\mathbb{C}}$
  Analytic Torsion}

In this subsection, as a warm-up,
we consider the Abelian case, i.e.\ $G=U(1)$, 
extended by the above $\RR_+$ gauge field to a $U(1)_{\CC}$-connection
on a complex line bundle $\mathcal{L}$, 
i.e.\ we study the determinant
\be
\Det_{\Omega^{1}(S^{1},\, \mathcal{L})}{\left(R^{-1}
    \left(i\frac{d}{d\theta} + \nu\right)\right)} 
\ee
of the operator 
\be
R^{-1} \left(i\frac{d}{d\theta} + \nu\right)
=R^{-1} \left(i\frac{d}{d\theta} + (\alpha + i\mu)\right)
\ee
with $\nu = \alpha + i\mu \in \CC$ representing a constant 
$U(1)_\CC$ gauge field, the metric dependence being encoded 
in the ``einbein'' $R^{-1}$ corresponding to the metric 
$g_{S^1} = R^2 d\theta \otimes d\theta$. 

The spectrum of this operator is simply $\{R^{-1}(n+\nu)\}$ with $n\in\ZZ$.
In particular, note that for $\mu \neq 0$ this operator has no zero
modes for any value of (the $U(1)$-part of the connection) $\alpha$,
including $\alpha=0$. As a consequence, we expect the Analytic Torsion
to be metric independent for $\mu \neq 0$. On the other hand, 
we can think of $\mu$ as encoding the mass parameter, and from that 
point of view we would expect the Analytic Torsion to be metric 
dependent. 

To see how this works out, note that
the spectral $\zeta$-function $\widehat{\zeta} (s, \nu)$ for this Dirac type
operator can, somewhat as in Example \ref{mass-s1} above, be written in terms of the 
Hurwitz $\zeta$-function $\zeta(s;a) = \sum_{n\geq 0} (n+a)^{-s}$ 
as
\be
\widehat{\zeta} (s, \nu)=( R)^{s} \Big[\zeta(s,
\nu)+ (-1)^{-s} \zeta(s, 1-\nu) \Big] 
\ee
The $R$-dependent term after taking the derivative evaluated at $s=0$ gives
(with $\zeta(0;a) = 1/2 -a$)
\be
\ln (R) \Big[\zeta(0, \nu)+  \zeta(0, 1-\nu)
\Big]= \ln (R) \Big[\frac{1}{2}-\nu
-\frac{1}{2}+ \nu \Big]=0 
\ee
and so we see explicitly that, as anticipated, 
there is no metric dependence with $\nu$, and hence $\mu$, regarded
as a connection. Then, using again $\zeta^\prime(0,a) = 
\ln \Gamma(a) -\ln (2\pi)/2$ and $\Gamma(a)\Gamma(1-a) = \pi/\sin(\pi a)$, 
the determinant evaluates to 
\be
\Det_{\Omega^{1}(S^{1},\, \mathcal{L})}{\left(R^{-1}
    \left(i\frac{d}{d\theta} + \nu\right)\right)} 
= \ex{-i\pi/2} \ex{i\pi \nu}\; 2 \sin \pi \nu\;\;.
\ee

If, however, $\mu$ is simply used to encode the (metric independent) 
mass term $m$, as in 
\be
R^{-1}i \frac{d}{d\theta}  + im = R^{-1}\left(i \frac{d}{d\theta} + i \mu\right)
\ee
with $\mu = mR$, then the situation is different. As explained
previously one may consider
$\mu d\theta$ to be a connection on a real line bundle $\mathbb{L}$. While 
we have shown that the overall $R^{-1}$ prefactor does not
enter into the final answer, the result depends on $R$ via the 
dependence of $\mu$ on $R$.
Indeed, then one has 
\be
\Det_{\Omega^{1}(S^{1},\,
  \mathbb{L})}{\left(R^{-1}\left(id/d \theta + i\mu \right)\right)} =
\ex{-i\pi/2}\, \ex{(- \pi \mu)} \, 2 
\sin{\left(i\pi \mu\right)} 
\ee
Clearly, for $\mu$ a metric independent connection, 
this result does not depend on the metric. On the other hand, 
for $\mu = mR$, equally clearly the result \textit{is} metric dependent, 
and for small mass we get the characteristic 
dependence 
\be
\Det_{\Omega^{1}(S^{1})}{\left(id/d\theta+ imR\right)} \longrightarrow
 2\pi mR 
\ee
on the metric. This resolves the issue raised at the beginning of this section
regarding the metric (in-) dependence of this quantity in a satisfactory (all is
as it should be) way.

\subsubsection{Ray-Singer Torsion for $G \times \mathbb{R}_{+}$ Connections}
\label{RSTM}

We now return to the general problem of determining
\eqref{taupm}. Without loss of generality, on $S^1$ we can assume that
$\phi$ is constant (we will later 
on re-derive the result from the path integral point of view without this 
assumption). We expand in terms of Fourier modes to obtain 
\be
\begin{aligned}
\tau_{S^{1}}(\phi + \mu,
\mathbb{E}\otimes \mathbb{L}) &= \Det_{\Omega^{1}(S^{1},
  \mathbb{E}\otimes \mathbb{L})}{\left(R^{-1}
    (iD_{\phi} + i \mu) \right)} \\
  & = \prod_{\lambda}\prod_{n \in \mathbb{Z}} R^{-1} (n+ \lambda(\phi) + i \mu)
\end{aligned}
\ee
where the $\lambda(\phi)$ are the real eigenvalues of the action of
$i\phi$ in the representation defined by $\mathbb{E}$. Note that, by a
reordering of the integers $n$ one
can take $0 \leq \lambda(\phi) < 1$ (as long as $\mu \neq 0$). As we
need to make sense of this first order operator we make use of the
Hurwitz $\zeta$-function regularisation, which amounts to setting
\be
\Det_{\Omega^{1}(S^{1},\mathbb{E}\otimes \mathbb{L})}{\left(iD_{\phi} + i
    \mu\right)}= \prod_{\lambda} \exp{\left(- \widehat{\zeta}\, '(0,
    \lambda(\phi)+ i \mu)  \right)}
\ee
where
\begin{align}
\widehat{\zeta}(s, \lambda(\phi)+ i \mu) &= \sum_{n \in \mathbb{Z}}
\frac{R^{s}}{(n+ \lambda(\phi) + i \mu)^{s}}
\nonumber\\
&= \sum_{n \geq 0} \frac{R^{s}}{(n+ \lambda(\phi) + i \mu)^{s}} +
(-1)^{-s} \sum_{n \geq 0} \frac{R^{s}}{(n + 1 - \lambda(\phi) - i
  \mu)^{s}} 
\nonumber\\
&= R^{s}\left[\zeta(s,\lambda(\phi)+ i \mu) + \ex{-i\pi s} \zeta(s, 1-
  \lambda(\phi)- i \mu)  \right]
\end{align}
with the $\zeta$-functions on the right hand side being the usual
Hurwitz $\zeta$-function and we have taken $-1 = \exp{(i\pi)}$. Note
that none of the derivatives of the Hurwitz $\zeta$-functions are
singular at $s=0$.

As in the previous Abelian example, 
with this regularisation there is no $R$ dependence as the only source
for such a dependence comes from differentiating the overall $R^{s}$
term which gives, at $s=0$,
\be
\ln{R}\, . \, [\zeta(0,\lambda(\phi)+ i \mu) + \zeta(0, 1-
  \lambda(\phi)- i \mu)] = 0
  \ee
  on using $\zeta(0,a) = 1/2 -a$. Consequently, with $z=\lambda(\phi) + i\mu$
and $\zeta^\prime(0,a) = \ln \Gamma(a) - \ln (2\pi)/2$, one finds 
\begin{align}
\widehat{\zeta}\, '(0,z) &=  \zeta'(0,z) + \zeta'(0, 1-z) - i\pi
\zeta(0, 1-z) \nonumber\\
  &= \ln{\Gamma(z)\Gamma(1-z)} - \ln{2\pi} + i\frac{\pi}{2} -i\pi z
\end{align}
Hence
\begin{align}
\Det_{\Omega^{1}(S^{1},\mathbb{E}\otimes \mathbb{L})}{\left(iD_{\phi} + i
    \mu\right)} &= \prod_{\lambda} -2i\sin{\pi (\lambda(\phi) +
  i\mu)}. \ex{i\pi (\lambda(\phi) + i\mu)} \nonumber\\
&= \Det_{E\otimes L}{\left(1- \ex{-2\pi\mu}. \rho(\ex{2\pi\phi})^{-1}\right)}
\label{detphimu}
\end{align}
where $\rho(\ex{2\pi\phi})$ and $\ex{2\pi\mu}=h$ are the holonomies of
the connections $\phi$ and $\mu$ along the
circle respectively. More generally, for $\phi = \phi(s)$ and $\mu = \mu(s) \neq 0$ not necessarily 
constant we will set 
\be
 g = \mathrm{P} \ex{\int_{0}^{2\pi} \phi(s) ds} \quad,\quad
h = \ex{\int_{0}^{2\pi} \mu (s) ds}\;\;.
\label{ghdef}
\ee
We have therefore established that the Ray-Singer Torsion is
\be
\tau_{S^{1}}(\phi + \mu, \mathbb{E} \otimes \mathbb{L}) = \Det_{E\otimes L}{\left(1-
    \ex{-2\pi\mu}. \rho(\ex{2\pi\phi})^{-1}\right)} \label{RST-AT}
\ee
One should note that with all our assumptions the determinant that
appears here has no zeros for $\mu > 0$ and as there are no zero modes
one has that the Analytic Torsion and Ray-Singer Torsion agree
\be
\widehat{\tau}_{S^{1}}(\phi + \mu, \mathbb{E} \otimes \mathbb{L}) =
\tau_{S^{1}}(\phi + \mu, \mathbb{E} \otimes \mathbb{L})
\ee

\begin{example}{Let $g\in SU(2)$ and $\rho$ be the fundamental
      representation. There is just one type of conjugacy class and one has
      \[
      g = k^{-1}.\left( \begin{array}{cc}
                          \ex{i \alpha} & 0\\
                          0 & \ex{-i \alpha}
                              \end{array}\right).k 
      \]
so that the Ray-Singer Torsion
for $SU(2)\times \mathbb{R}_{+}$ is
\[
\widehat{\tau}_{S^{1}}(\phi + \mu, \mathbb{C}^{2}\otimes
L ) = 1 - 2h^{-1}\cos{\alpha} 
+ h^{-2}= (1-h^{-1})^2 + 4 h^{-1}\sin^2\alpha/2
\]
Clearly, this does not vanish for any positive $h\neq 1$ (i.e.\ for any 
real non-zero $\mu(x)$), reflecting the fact that 
the zero-modes have been lifted by the mass term.}
\end{example}
In the previous example $h$ was introduced to lift the zero modes and
it did just that. Surprisingly even for non-compact groups, for which not
all of our assumptions (like the compatibility of the connection with a 
positive-definite scalar product on the fibres) need to be satisfied, it may
happen that the introduction of a mass term eliminates the zero modes, 
though the situation may be
considerably more complicated. An example of interest is a calculation
arising in the context of JT gravity in 2 dimensions \cite{Stanford-Witten}.
Amongst other things the authors determine the
Reidemeister Torsion on the circle for the group $SL(2,
\mathbb{R})$. In the following example we determine the Ray-Singer
Torsion and establish equivalence with the result in \cite{Stanford-Witten}.

\begin{example}{Let $g \in PSL(2, \mathbb{R})$ then there are three
      distinct types of conjugacy classes depending on whether the trace is
      less than 2 (elliptic), equal to 2 (parabolic) 
or greater than 2 (hyperbolic), 
      \[
     g =  \left( \begin{array}{cc}
                \cos{\theta}   & \sin{\theta} \\
                   -\sin{\theta} & \cos{\theta}
                 \end{array}
               \right), \;\;\; \left( \begin{array}{cc}
                  1 & \pm 1\\
                  0 & 1
                 \end{array}
               \right), \;\;\; \left( \begin{array}{cc}
                  b & 0\\
                  0 & b^{-1}
                 \end{array}
               \right)
               \]
with $\theta \in [0,\pi)$ and $b\geq 1$.
The corresponding Ray-Singer Torsion in the
adjoint representation $\mathfrak{psl}(2, \mathbb{R}) \simeq \mathbb{R}^{3}$
is
               \[
               \tau_{S^{1}}(\phi + \mu, 
               \mathbb{R}^{3}\otimes L) = \left\{ \begin{array}{l}
                  (1 - h^{-1})^{3} +(1-h^{-1})h^{-1} 4\sin^{2}{\theta} \\
                                                    (1-h^{-1})^{3} \\
                              (1 - h^{-1})^{3} -(1-h^{-1}) h^{-1}(b - b^{-1})^{2}
                                                    \end{array} \right.
                                                  \]
respectively. In all three
cases there are zero modes for $\mu=0$, i.e.\ for $h=1$. 
However, for $ h \neq 1$, the torsion is clearly 
regularised (that is it is not zero due to zero modes) in the elliptic and 
parabolic case. In the hyperbolic case, there are zero modes for 
$h = b^{\pm 2}$, but nevertheless 
the torsion is regularised for all other values, in particular for $h \in 
(b^{-2},b^{2})$, and one can then look at the limit $h \ra 1$ with impunity.

At the identity matrix ($\theta = 0$ 
or $b= 1$) and in the parabolic case,  
the above expressions have a zero of order 3 in $\mu$. In all other cases, 
one has zeros of order one. 
We can now readily determine the Ray-Singer Torsion in each case. 
To that end one first passes to the
massive Ray-Singer Torsion in
factorised form (\ref{RST-fact}) and then takes the limit as in
(\ref{Det-project}) with the appropriate power of $\mu$. One finds
\[
               \widehat{\tau}_{S^{1}}(\phi, 
               \mathbb{R}^{3}) = \left\{ \begin{array}{l}
                  4\sin^{2}{\theta},\quad \quad 1 \; {\rm if}\; \theta =
0 \\ 
                  1 \\
                  (b - b^{-1})^{2}, \quad 1 \; {\rm if}\; b=1
                  \end{array} \right.
                                                  \]
This result agrees (for the hyperbolic conjugacy class) with the Reidemeister
Torsion calculation in \cite{Stanford-Witten} (cf.\ their equation (3.42)
with $b = \exp{a/2}$). The `accidental' zero modes occur in every
finite dimensional representation of $PSL(2, \mathbb{R})$ as then the holonomy for the
hyperbolic elements times the $\mathbb{R}_{+}$ holonomy will be 
\[
h \left(\begin{array}{cccc}
b^{r}& 0 & \dots & 0\\
          0 & b^{r-2} & \dots & 0\\
          \dots & \dots & \dots & \dots\\
          \dots & \dots & b^{2-r} & 0\\
          0 & \dots & 0 & b^{-r}
\end{array} \right) 
\]
which has eigenvalue 1 whenever $h = b^{r}, b^{(r-2)},
\dots,b^{(2-r)}, b^{-r} $.
}
\end{example}

\subsubsection{Passing from $G \times \mathbb{R}_{+}$ to $G$
  Ray-Singer Torsion}\label{MDA}

We briefly review the relationship between zero modes and reducibility
of the holonomy $g$ (the unit eigenvalues of $\rho(g)$). Let $\psi \in
\Gamma(S^{1}, \mathbb{E})$ be an eigenvector with zero eigenvalue then
\be
D_{\phi}\psi =0
\ee
The solution is
\be
\psi(\theta) = \mathrm{P}\exp{\left(\int_{0}^{\theta}
    \rho(\phi(s)) ds\right)}.\psi(0)
\ee
However, periodicity requires that $\psi(2\pi) = \psi(0)$ or that
\be
(1-\rho(g))\psi(0)=0 \label{g-inv}
\ee
Clearly with $\mu=0$ the zero modes of the Laplacian translate into
the zeros of the determinant on the right hand side of
(\ref{RST-AT}). 

So in order to isolate the contribution of the zero modes in general, 
split the representation space 
$E$ as $E = E_{\parallel} \oplus
E_{\perp}$ where the vectors in $E_{\parallel}$ are
invariant under the action of $g= \ex{2\pi \phi}$, that is they
satisfy (\ref{g-inv}). Then according to
(\ref{Det-project}) the Analytic
Torsion is obtained as the limit
\begin{align}
\tau_{S^{1}}(\phi, \mathbb{E})& =  \lim_{m \rightarrow 0}
m^{-\dim{E_{\parallel}}} 
.\Det_{E}{\left(1 
    -h^{-1}\rho(g)^{-1}\right) }\nonumber\\
&=(2\pi R)^{\dim{E_{\parallel}}} \Det_{E_{\perp}}{\left(1 
    -\rho(g)^{-1} \right) } \label{rst-gen1}
\end{align} 
This is not quite the Ray-Singer Torsion as we still need to multiply
by the metric factor $\rho_{H}(R)$ which we will do below.

\begin{example}\label{RST-Ad}{The situation we most often come across in gauge theory
      is where $\mathbb{E}$ is the $ \ad{P}$ bundle with fibre $\lg$
      the Lie algebra of the compact gauge group $G$. Split the
      complexified Lie algebra as 
      $\lg_{\mathbb{C}}= \lt_{\mathbb{C}} \oplus
      \lk_{\mathbb{C}}$ where $\lt$ is the Cartan subalgebra. 
      Providing $\phi$ is a regular element 
      of $\lg$ so that its centraliser can be taken to be $\lt$, write
      (somewhat incorrectly)
      $E = \lt \oplus \lk$ where $\lt =
      E_{\parallel}$ and $\lk =
      E_{\perp}$ so that in this case from
      (\ref{rst-gen1}) we have
      \begin{align}
\tau_{S^{1}}(\phi, \lg) 
&=\lim_{m \rightarrow 0} (m)^{-\dim{\lt}}
.\Det_{\lg}{\left(1- h^{-1}\Ad{(g)}^{-1} 
   \right) }\nonumber\\
& = (2\pi R)^{\dim{\lt}}  \Det_{\lk}{\left(1-\Ad{(g)}\right) } \nonumber
\end{align}
This will go over to the Ray-Singer Torsion that we found in
\cite{BT-CS} or as Reidemeister torsion as found by Freed 
\cite{Freed} or Witten \cite{Witten-2d1} once we have included the
$\rho_{H}(R)$ contribution (and we do this in Section \ref{MDA} below). 
Note that if $\phi$ is not regular, its
centraliser (and thus $E_{\parallel}$) is larger but we
may still apply (\ref{rst-gen1}) to get a sensible answer. For example,
if $\phi=0$, then $E_{\parallel}= \lg$, thus $E_\perp$ is
simply absent, and
\[
\tau_{S^{1}}(\phi=0)= (2\pi R)^{\dim{\lg}}\nonumber
\]
}
\end{example}

In summary,we have seen that 
the introduction of a mass has not only lifted the zero modes: it has also
allowed us to follow the
non-trivial cohomology in case the connection does not lead to acyclic
cohomology and to pick out precisely the terms where that cohomology has been
projected out. One may, if one wishes, retain and 
follow the $\mu$ (or $h$) dependence and we
will do so further along.

Within the context of the massive Ray-Singer Torsion we must reconsider the
issue of metric dependence given the twin role played by $\mu$.
If one considers $\mu$ to be an $\mathbb{R}_{+}$ flat connection then
there is nothing to do. Indeed by (\ref{RST-AT}) the Analytic Torsion for the $G \times
\mathbb{R}_{+}$ connection is the same as the Ray-Singer Torsion and
is respectably metric independent.

If, on the other hand, $\mu=mR$ is simply a mass regulator then to arrive
at the Ray-Singer Torsion for a $G$ connection there are two passes that we need to
make. Firstly one should take the $m \rightarrow 0$ limit as in
(\ref{Det-project}) of the Analytic Torsion with mass and then one 
must multiply the result by
$\rho_{H}$ to arrive at the metric-independent 
Ray-Singer Torsion. We will now see
that the only possible natural combination of $m$ and $R$ arises in
this way, namely the combination $\mu = mR$, in terms of which 
(as we have just seen) there is indeed no metric dependence.

By the normalisation convention \eqref{s1norm}, we find a factor 
of $(2\pi R)^{-1}$ for each orthonormal basis vector $e_i$ of
the vector space $E_{\parallel}$. Therefore, 
\be
\rho_{H}(R) = (2\pi R)^{- \dim{E_{\parallel}}} 
\ee
while from (\ref{rst-gen1}) we are asked to divide by
$m^{\dim{E_{\parallel}}}$. The combination then is to take
\be
\begin{aligned}
\widehat{\tau}_{S^{1}}(\phi, \mathbb{E}) &= \lim_{m\rightarrow 0} \rho_{H}(
mR)\tau_{S^{1}}(\phi + \mu, \mathbb{E}) = \lim_{\mu \rightarrow 0} \rho_{H}(
\mu)\tau_{S^{1}}(\phi + \mu, \mathbb{E})\\
& =  \Det_{E_{\perp}}{\left(1 
    -\rho(g)^{-1} \right) } \label{rst-gen2}
\end{aligned}
\ee
From this
perspective, dividing by a suitable power of the constant $\mathbb{R}_{+}$ 
connection $\mu=mR$ rather than of the mass ensures that at no point in the 
calculation any metric dependence arises. 

In particular the Ray-Singer Torsion for the adjoint representation
Example \ref{RST-Ad} is (for $\phi$ regular)
\be
\widehat{\tau}_{S^{1}}(\phi, \lg)= \Det_{\lk}{\left(1-\Ad{(g)}\right) }
\ee
agreeing with the results obtained in \cite{BT-CS} and the
Reidemeister form in \cite{Freed, Witten-2d1}.

Before concluding this section we observe that as there are no zero
modes on $\mathbb{E}_{\perp}$ that
\be
\widehat{\tau}_{S^{1}}(\phi, \mathbb{E}_{\perp}) = \tau_{S^{1}}(\phi, \mathbb{E}_{\perp})
\ee
and that essentially the prescription for dealing with the zero modes
says that the Ray-Singer Torsion on $\mathbb{E}$ is the Analytic
Torsion on $\mathbb{E}_{\perp}$ so we end up with a cycle of
equivalences
\be
\widehat{\tau}_{S^{1}}(\phi, \mathbb{E}) =\widehat{\tau}_{S^{1}}(\phi, \mathbb{E}_{\perp}) = \tau_{S^{1}}(\phi,
\mathbb{E}_{\perp})
\ee
This seems to imply that the prescription of Ray and Singer for
calculating the Ray-Singer Torsion on a bundle $\mathbb{E}$ is
essentially to calculate the Analytic Torsion on the bundle $\mathbb{E}_{\perp}$.

\subsubsection{Path Integral Representation}\label{MRSTS1}

It is difficult to see, in general, how to construct a first order
action with mass which has a quantum gauge symmetry without introducing
more fields and such that it reproduces the action (\ref{RST-Action})
as the mass goes to zero. Consequently, it is not clear how to
create the Schwarz type topological theory corresponding to the Massive
Ray-Singer Torsion  that we have defined. 

However, as we have already seen, on $S^1$ a mass term can be converted into 
a gauge field, and moreover on $S^1$ the gauge shift symmetry is not
an issue. As we will see later on, these favourable features are 
inherited by Schwarz-type gauge theories on certain fibrations 
over $S^1$, where the torsion on $S^{1}$ will continue to play a central
role. 

Given its importance, we therefore now turn to the torsion on $S^{1}$
from a path integral perspective. In particular we will evaluate
the path integral for the torsion of $S^1$ by a small but useful
variation of the method used in \cite{BT-CS}, as this new method
turns out to generalise in a rather straightforward way to more
general manifolds (such as mapping tori - see Section \ref{sec-map}).

The action that we use for the path integral to 
represent the determinants that 
we have already studied in Section \ref{RSTM} is (see for
example (3.9) in \cite{BT-CS})
\be
iS_{F} = \int_{S^{1}}R^{-1} \; \langle \overline{\eta} , (iD_{\phi}
 \pm i \mu)
\eta\rangle d\theta \label{SF}
\ee
where $\eta$ and $\overline{\eta}$ are Grassmann odd sections of $\mathbb{E}$
and $\mathbb{E}^{*}$ respectively. Because of its metric dependence, 
encoded in the ``einbein'' $R^{-1}$, this is not strictly speaking at first 
sight just a Grassmann-odd one-dimensional version of the Schwarz-type 
actions \eqref{RST-Action}, which are written in terms of differential forms 
and are thus explicitly metric-independent. Rather, written in this way
the action is more like a 1-dimensional Dirac action. Nevertheless, once
we have established the $R$-independence of the result (and we will do this
in detail below), we may just set $R=1$ and view the above action as 
nothing other than a Grassmann-odd Schwarz action
(cf.\ also the comment in Section \ref{subsubmiya} below).

Now let us turn to the path integral. 
If one uses a Fourier mode
expansion the measure is taken to be
\be
 D\eta D\overline{\eta} = \prod_{n}  d\eta_{n}
d\overline{\eta}_{n}
\ee
so that the integral over the exponential of (\ref{SF}) yields
$\Det_{\Omega^{1}(S^{1},\mathbb{E}\otimes \mathbb{L})}{\left(iD_{\phi} \pm i
    \mu\right)}$ and one could evaluate it with the $\zeta$-functions
of the previous section.

Our aim here is to give an evaluation of the corresponding partition
function
\be
Z[\phi \pm \mu, R] = \int D\eta D\overline{\eta}\;
\exp{\left(\int_{S^{1}}R^{-1} \; \langle \overline{\eta} , (iD_{\phi} 
 \pm i \mu)
\eta\rangle d\theta \right)}
\ee
where the Grassmann fields are periodic on
$S^{1}$. The usual normalisation of the path integral measure is so
that the one loop contribution has no coupling constant pre-factor. In
this case it would mean that we have the measure
\be
D(\eta/\sqrt{R}) D(\overline{\eta}'/\sqrt{R})
\ee
and a scaling $(\eta, \overline{\eta}) \rightarrow (\sqrt{R}\eta',
\sqrt{R}\overline{\eta}')$ would eliminate $R$ from both the action as
well as the measure. Rather more generally, if we start with the
measure which does not involve $R$, then any $R$-dependence that arises
can be eliminated by a renormalisation, depending on the
regularisation, of the log of the volume of the 
space. So we could always multiply the partition function by 
a term of the form
\be
\exp{\left(a + b\ln{R}\right)} \label{renorm}
\ee
This is quite analogous to the ``standard renormalisation'' ambiguities 
encountered 
e.g.\ in Yang-Mills theory on a Riemann surface \cite{Witten-2d1, BT-YM2}.
With the $\zeta$-function regularisation we see that there is
no need to renormalise with the volume $2\pi R$. For any other
regularisation the counter terms would need to be chosen so that the
metric independence is maintained. At that point the definition of
what one means by the Ray-Singer Torsion in terms of path integrals would
include the renormalisation prescription. With this understood we may set
$R=1$ for now and in so doing we have that the action \eqref{SF} 
is of the standard metric-independent 
Schwarz type (albeit with Grassmann-odd fields).

In the evaluation of this path integral we do not presume that $\phi$ or $\mu$ are time
independent.  We will evaluate this
partition function by using a slight, but useful, variant of
the trick used in \cite{BT-CS}. In order to do so we will have to
explicitly impose the conditions that the variables are periodic. To
do this one incorporates the Grassmann delta functions
\be
\delta\left(\eta(2\pi)-\eta(0) \right)\,
\delta\left(\overline{\eta}(2\pi)- \overline{\eta}(0) \right) \label{periodic-delta-1}
\ee
in the measure.

In order to simplify the evaluation of the path integral, 
we now perform the change of variables
\be
\eta = \ex{\mp\int_{0}^{\theta} \mu(s) ds }\, \rho(g(\theta))^{-1}\, . \, \eta',
\;\;\; \overline{\eta} 
\rightarrow \ex{\pm \int_{0}^{\theta} \mu(s) ds}\, \overline{\eta}'\,
. \, \rho(g(\theta))
\ee
where $g(\theta)$ solves the equation
\be
 g(\theta)D_{\phi}g(\theta)^{-1} =0 \;\;\; g(0)= 1, 
\ee
and $\rho$ indicates the representation defined by $\mathbb{E}$. 
Note that by \eqref{ghdef} we have 
\be
g(\theta = 2\pi) \equiv g  =
 \mathrm{P} \ex{\int_{0}^{2\pi} \phi(s) ds}
 \ee
The periodicity constraints (\ref{periodic-delta-1}) now go over to
\be
\delta\left(\eta'(0) - h^{ \mp 1}\, \rho(g)^{-1} .\, \eta'(2\pi)\right)\,
\delta\left( \overline{\eta}'(0)-  h^{\pm 1}\, \overline{\eta}'(2\pi) \, 
\rho(g)\right) \label{new-bc}
\ee
with $h = \exp{\left(\int_{0}^{2\pi} \mu (s) ds\right)}$ \eqref{ghdef}.

The point of these transformations is that in terms of the new fields
the action is free
\be
iS_{F} = \int_{S^{1}}d\theta \, \langle \overline{\eta}', i
\frac{\partial}{\partial \theta} \eta' \rangle \label{free-act}
\ee
while the measure goes over to $D\eta' D \overline{\eta}'$, however,
with the new boundary conditions (\ref{new-bc}). Next we discretise the
circle as in \cite{Singh-Steiner}. Replace the continuous variable
with the distinct points $i \in [0, N]$ with separating distance
between consecutive points being $2\pi/N$. Because of the
delta function `periodicity' conditions the path integral can be taken
to be over all $\eta'_{i}$ and
$\overline{\eta}'_{i}$ for $i= 0, \dots, N$. The idea is to evaluate the
finite dimensional discretised path integral and then take the $N
\rightarrow \infty$ limit. 

The choice for the discretised version of (\ref{free-act}), as
explained in \cite{Singh-Steiner}, is
\be
iS_{F} \rightarrow i \sum_{j=1}^{N}\, \langle \overline{\eta}'_{j},
(\eta'_{j}-\eta'_{j-1}) \rangle
\ee
Notice that $\overline{\eta}'_{0}$ makes no appearance in the action
so that  one can use the periodicity condition to remove it completely
from the path integral. $\eta'_{0}$ does appear in the action but is determined by
the boundary conditions. The
integral over the $\overline{\eta}'_{j}$ gives a set of delta 
functions. Integrating over $\overline{\eta}'_{1}$ means we can do the
$\eta'_{1}$ integral which sets it to $\eta'_{0}$. The
$\overline{\eta}'_{2}$ integral sets $\eta'_{2}$ to $\eta'_{1}$ which
is now $\eta'_{0}$ and so on until we get to the last integral over
$\overline{\eta}'_{N}$ which sets $\eta'_{N}$ to $\eta'_{N-1}
=\eta'_{0} $. The finally we end up with
\be
i^{N \dim{\mathbb{E}}}\int d\eta'_{0}\, 
  \delta((1- h^{\mp 1}\rho(g)^{ -1})\eta'_{0}) =  i^{N
    \dim{\mathbb{E}}}\Det_{\mathbb{E}}{\left(1-h^{\mp 1} \rho(g)^{
      -1}\right) }
\ee
We should take the $N \rightarrow \infty$ limit to arrive at the
continuum. Only the prefactor changes and that can be eliminated by a
finite renormalisation on appropriate choice of $a$ in (\ref{renorm}) 
(e.g.\ the regularisation used in
  \cite{BT-CS} does not give rise to the $i^{N
    \dim{\mathbb{E}}}$ prefactor).
That the answer depends only weakly on how one 
treats the circle (and is essentially independent of the discretisation)
is a manifestation of the topological nature of the
theory. 
Then we are formally left with
\be
Z[\phi \pm \mu, R]= \Det_{E \otimes L^{\pm 1}}{\left(1-h^{\mp 1}\rho(g)^{-1}
    \right) }\label{detE}
\ee
as the partition function. We have thus rederived the result \eqref{detphimu}
directly from the path integral.

\section{Massive Ray-Singer Torsion on $N \times S^{1}$}\label{mrstns1}

Here we draw together several of the ingredients that have, separately,
appeared in the previous section: the field theoretic realisation
of the Ray-Singer Torsion, the massive deformation of the Ray-Singer
Torsion, and the interpretation of the mass term (on $S^1$) as a flat
background gauge field. We combine these by coupling the standard
Schwarz-type action \eqref{RST-Action} for the Ray-Singer Torsion to a
flat $\mathbb{R}_+$ gauge field, and then studying situations in 
which this gauge theory can indeed be regarded as a massive deformation
(and regularisation) of the original theory. Manifolds on which this is 
the case include $M=N\times S^1$ (studied in this section), and mapping
tori $M=N_f$ (which we will look at in Section \ref{sec-map}). We 
show that on such manifolds a purely algebraic gauge condition is 
available which greatly simplifies the evaluation of the partition function
and (essentially) reduces it to the calculation of the massive 
Ray-Singer Torsion on $S^1$, already discussed in detail in the previous
section. 

The field theory approach allows us to quantise in a
way that from the outset does not depend on the details of the product
metric that can be put on the manifold and allow us to obtain the
product formula (that is also derived through traditional means in the
appendix). As a side benefit we also obtain a deeper understanding of
why in the Abelianisation programme a connection that does not appear
to be flat appears as the argument of the Ray-Singer Torsion. We
include a derivation of the Schwarz type field theories directly from a path
integral that involves only Laplacians (no Dirac type operators) that
obviously represent the torsion.

\subsection{The Classical Action and its Symmetries}

As just explained, the action of interest is now 
the action \eqref{RST-Action} for
the fields $\mathscr{B}$ and $\mathscr{C}$ coupled to a flat background
gauge field $A$, but with $A$ now extended to
include an $\mathbb{R}_{+}$ connection. Thus the action is
\be
S_{M}(A)= \int_{M} \langle \mathscr{B},  d_{A}
\mathscr{C}\rangle \label{RST-action}
\ee
for the fields with values in $\mathbb{E} \otimes \mathbb{L}^{\pm 1}$ with
$\mathbb{E} $ a vector bundle and
$\mathbb{L}$ a real line bundle over $M$. Specifically we choose 
$\mathscr{B}\in
\Omega^{p}(M, \mathbb{E} \otimes \mathbb{L}) $, $\mathscr{C}  \in
\Omega^{n-p-1}(M, \mathbb{E}\otimes \mathbb{L}^{-1})$, for $0 \leq p
\leq n-1$ where $\dim_{\mathbb{R}}{M}=n$. 
Moreover, $A$ is now a flat $G \times \mathbb{R}_{+}$ connection
for some (compact) gauge group $G$, which splits into a $\lg \oplus
\mathbb{R}$ part as 
\be
A= A^{\lg} + A^{\mathbb{R}} \;\;,
\ee
and the action of $\mathbb{R}_{+}$ on the fields is
\be
A \rightarrow A + d\alpha, \quad \mathscr{B} \rightarrow
\ex{\alpha}.\mathscr{B}, \quad \mathscr{C} 
\rightarrow \ex{-\alpha}.\mathscr{C}, \quad \alpha : M \rightarrow \mathbb{R}
\ee

Flatness of the extended connection 
$A$ implies that the action then still has the shift symmetry
\eqref{qsym}, 
\be
\delta \mathscr{B}  = d_{A}\Sigma \quad,\quad
\delta \mathscr{C} = d_{A} \Lambda
\label{RST-symm}
\ee

Thus, provided that one has  a non-trivial flat connection 
$A=A^\lg + A^{\RR}$ (both parts need to be flat separately), 
the action has the required shift gauge invariance. Moreover, 
by construction, the (covariantly) gauge fixed 
partition function will formally calculate the Ray-Singer Torsion 
$\tau_{S^1}(M,A^\lg + A^\RR,\EE \ot\mathbb{L})$.

However, while on the circle
$S^1$, the non-trivial flat connection $A^{\RR}=\mu d\theta$ was enough 
to regularise the theory by providing a mass term that lifts the zero
modes of the kinetic term, on more general manifolds
a non-trivial flat connection $A^{\RR}$ will in general 
not be sufficient to regularise the theory. 

We will not try to give a complete answer to the question, under which
conditions the coupling to a flat $\mathbb{R}_+$-connection is 
sufficient to ensure that all the zero modes of the theory are lifted. 
However, we can easily identify two necessary conditions, and we can 
then also identify situations in which we see (by explicit calculation)
that these conditions are also sufficient:
\begin{enumerate}
\item First of all, there are no global topological issues associated
with an $\mathbb{R}_+$-bundle (with its contractible structure group). 
Thus we can globally identify the gauge field $A^\RR$ with a 1-form on
$M$. Flatness of $A^\RR$ is then simply the statement that $A^\RR$ is closed, 
$dA^\RR=0$.
In order for such an $A^\RR$ to not be globally gauge equivalent to 
$A^\RR=0$ (which would not be helpful), $A^\RR$ must be closed but not 
exact. Thus our first necessary condition is that 
\be
H^1(M,\RR) \neq 0\;\;.
\ee
\item Secondly, 
we would only expect $A^\RR$ to act like a mass term that can lift 
all the zero modes if it is nowhere vanishing. By duality (using any 
metric $g_M$ on $M$), this requires $M$ to admit a nowhere vanishing
vector field, and a necessary condition for this is
that the Euler characteristic of $M$ vanishes, 
\be
e(M) = 0 \;\;.
\ee
\end{enumerate}

In the following we will look at two classes of manifolds where
these two conditions are satisfied, namely product manifolds
$M=N\times S^1$ (in this section) and finite order mapping tori $M=N_f$, which
are non-trivial fibrations over $S^1$ (in Section \ref{sec-map}
below). In both cases, one can readily identify a non-vanishing 
vector field ($\del_\theta$ or $\del_t$ respectively) or appeal to the multiplicativity
of the Euler characteristic for fibrations to conclude 
\be
M= N\times S^1 \quad \text{or}\quad  M = N_f \quad\Ra\quad e(M)=0\;\;.
\ee
Moreover, in both cases, $d\theta$ or $dt$ respectively represents a non-trivial 
element of $H^1(M,\RR)$.

\subsection{Decomposition of Fields on $N\times S^1$}

We now consider the $G \times \RR_+$ 
theory described by the classical action 
\eqref{RST-action} on a manifold of the form $M=N\times S^1$. 
Given the product structure of the manifold, it will be convenient 
to decompose all the fields and gauge parameters accordingly. 

\begin{itemize}
\item The Connection and the Covariant Derivative

On $M = N \times
S^{1}$ we can further split the connection 
$A= A^{\lg} + A^{\mathbb{R}}$ as 
\be
A^{\lg} = A_{N}^{\lg}+ \phi d\theta\quad, \quad A^{\mathbb{R}} =
A_{N}^{\mathbb{R}}+ \mu d\theta\quad, \quad A = A_N + (\phi +\mu)d\theta
\ee
where $\phi \in \lg$ is anti-Hermitian and $\mu$ is real. 
Correspondingly, the covariant derivative splits as 
\be
d_A = d_N + A_N + d\theta (\del_\theta + \phi+\mu) 
\equiv d_{A_N} + d\theta D_{\phi+\mu}\;\;.
\ee
The operator $D_{\phi+\mu}$ will appear repeatedly in the following, and 
we will therefore abbreviate it to
\be
D \equiv D_{\phi+\mu}\;\;.
\ee

\item The Laplace Operator 

Also the massive Laplacian with respect to the
product metric $g_{M}=g_N + R^2 d\theta \otimes d\theta$ 
splits neatly as
\be
\Delta^{M}_{A} + m^{2}= \Delta^{N}_{A} + R^{-2}( i D_{\phi+\mu})(iD_{\phi-\mu})
\ee
where we have once again used the identification $\mu = mR$ 
\eqref{muRm}.
\item The Fields $\mathscr{B}$ and $\mathscr{C}$ 

We can also decompose the fields $\mathscr{B}$ and $\mathscr{C}$ 
on $M = N \times S^{1}$ as
\be
\mathscr{B}^{p} =  B^{p} +  B^{p-1} d\theta, \quad \quad \;\; \mathscr{C}^{n-p-1} =
C^{n-p-1}  + C^{n-p-2} d\theta
\ee
The fields that appear on the right hand side of these equations are all
horizontal
\be
\iota_{\xi}B^{p}=0, \quad \iota_{\xi}B^{p-1}=0, \quad
\iota_{\xi}C^{n-p-1}=0, \quad  \iota_{\xi}C^{n-p-2}=0
\ee
with respect to the $S^{1}$ vector
field $\xi$ (normalised so that $\iota_{\xi} d\theta =1$). This means
that they are forms of type $\Omega^{\bullet}(N, \mathbb{E}\otimes
\mathbb{L}^{\pm 1} ) \otimes \Omega^{0}(S^{1})$.

\item The Action 

With respect to this decomposition, the action \eqref{RST-action}
takes the form 
\be
S_{M}(A)= (-1)^p\int_{M} 
d\theta 
\langle 
B^p D C^{q} - B^{p-1}d_{A_N}C^{q} 
+ (-1)^{q}B^pd_{A_N}C^{q-1}  
\rangle \label{RST-action-2}
\ee
with $q=n-p-1$.

\item Decomposition of all other Fields

  All fields $\Phi$ other than $\mathscr{B}$ and $\mathscr{C}$ will be
  decomposed as follows
  \be
  \Phi^{k} = \varphi^{k} + \Phi_{k-1}\, d\theta 
  \ee
Capital letter for the starting field $\Phi^{k}$ of degree $k$
(exhibited as a
superscript), small letter for the component $\varphi^{k}$
which is a $k$-form on $N$ (exhibited as a
superscript) and a capital letter for the component of the form of
mixed degree $ \Phi_{k-1}$ which is a $k-1$-form on $N$ (exhibited as a
subscript).

\item The Shift Symmetry Parameters

With the previous notation in hand, the 
parameters $\Lambda \in
\Omega^{n-p-2}(M, \mathbb{E} \otimes \mathbb{L}^{-1})$ and $\Sigma \in
\Omega^{p-1}(M, \mathbb{E} \otimes \mathbb{L})$ 
appearing in the shift symmetry (\ref{RST-symm}) 
can also be decomposed
\be
\Sigma^{p-1} = \sigma^{p-1} +
\Sigma_{p-2}\, d\theta, \quad \Lambda^{q-1} = \lambda^{q-1} +
\Lambda_{q-2}\, d\theta
\ee
so that the symmetry reads
\begin{align}
 & \delta B^{p} = d_{A}^{N} \sigma^{p-1}, \quad 
\delta B^{p-1} = d_{A}^{N} \Sigma_{p-2} + (-1)^{p-1} D \sigma^{p-1}\nonumber\\
 & \delta C^{q} = d_{A}^{N} \lambda^{q-1}, \quad \delta C^{q-1} = d_{A}^{N}
\Lambda_{q-2} + (-1)^{q-1} D \lambda^{q-1}
\label{deltaBC}
\end{align}
In general there are also ghost for ghosts coming from the fact that
the symmetry (\ref{RST-symm}) is `reducible' namely, as $d_{A}^{2}=0$ it does not change
under 
\be
\Sigma^{p-1} \rightarrow \Sigma^{p-1} + d_{A}\Sigma^{p-2}, \quad \Lambda^{q-1}
\rightarrow \Lambda^{q-1}  + d_{A}\Lambda^{q-2}
\ee
and so on. Under the decomposition
\be
\Sigma^{p-2} = \sigma^{p-2} + \Sigma_{p-3}\, d\theta,
\quad \Lambda^{q-2} = \lambda^{q-2}  + 
\Lambda_{q-3}\,  d\theta
\ee
we have
\be
\sigma^{p-1} \rightarrow \sigma^{p-1} + d_{A}^{N} \sigma^{p-2}, \quad
\Sigma_{p-2} \rightarrow \Sigma_{p-2} + d_{A}^{N} \Sigma_{p-3} +
(-1)^{p} D \sigma^{p-2}\label{second-symm}
\ee
and a similar expression for $\Lambda$. One keeps
getting gauge symmetries of gauge symmetries until the last variation
is of the form $d_{A}\Sigma^{0}$ and, as it is a zero-form, in the
above notation one has $\Sigma^{0}= \sigma^{0}$. For $B^{p}$ there is a total nesting
of $p$ symmetries (and hence $p$ parameters) while for $C^{q}$ there is a total nesting
of $q$ symmetries.
\end{itemize}

\subsection{Algebraic Gauge Fixing Conditions}
\label{subagf}

As mentioned before, 
on any manifold $M$ one can gauge fix the shift symmetries
\eqref{qsym} or \eqref{RST-symm} of the action 
by imposing the usual covariant gauge conditions 
\eqref{cov-gf}, 
\be
d_{A}*\mathscr{B} =0, \quad d_{A}*\mathscr{C} =0
\label{cov-gf2}
\ee
on the fields $\mathscr{B}$ and $\mathscr{C}$. 

However, it turns out that 
with the extra structure afforded to us by having a preferred
direction and an
$\mathbb{R}_{+}$ connection (mass term) on $M= N \times S^{1}$, an
algebraic gauge condition is available: namely one may simply
set the $S^1$-components $B^{p-1}$ and $C^{q-1}$ of the fields
$\mathscr{B}$ and $\mathscr{C}$ to zero, 
\be
B^{p-1}= 0, \quad \mathrm{and} \quad C^{q-1}=0 \label{gfixing}
\ee
Before turning to a proof of this statement, let us comment on some
implications of this result:
\begin{itemize}
\item The great advantage of such a gauge
choice is that the action \eqref{RST-action-2} 
simplifies immensely; indeed evidently it reduces to the simple
action
\be
S_{M}(A) = (-1)^{p}\int_{M}d\theta \langle B^{p},
  DC^{q} \rangle  \label{RST-action-1}
\ee
\item 
The same argument that establishes \eqref{gfixing} can also be used to
set the $S^1$-components of the gauge (and gauge for gauge) parameters, 
namely all the $\Sigma_{k}$ and 
$\Lambda_{k}$, to zero. This also leads to a tremendous simplification of the full quantum action (see the examples and the discussion below), 
and of the calculation of the partition function.
\item Moreover, the gauge condition \eqref{gfixing} is manifestly 
metric independent, something that should not be underrated in the 
context of topological field theories.
\end{itemize}

Let us now turn to the gauge conditions \eqref{gfixing} themselves.
These conditions are of course in some sense the higher degree form
analogues of the temporal gauge $A_{0}=0$ in QED or QCD. However, it 
is important to also keep in mind the differences: for a standard 
gauge field $A$ with gauge transformation 
$A\ra A^g = g^{-1}Ag + g^{-1}dg$, this temporal gauge
is available on any manifold of the form $M = N \times \RR$. However, 
in general, on a manifold of the form $M=N\times S^1$, this gauge condition
cannot be imposed (the obstruction being the possible non-trivial holonomy
of $A$ along the $S^1$), and the ``time''-dependent gauge transformations
can only be used to impose the weaker condition $\del_0 A_0=0$. 

What we will now show is that for a field $\mathscr{B}$ transforming
with what we have called the shift symmetry $\delta\mathscr{B} =
d_A\Sigma$ (rather than like a connection), we \textit{can} impose
such a temporal gauge even on $M=N\times S^1$ , provided that the
$S^1$-component $D$ of the operator $d_A$ is invertible. In the
case at hand this is accomplished by the mass term or $\RR_+$-connection - 
after all, this was the purpose of introducing the mass term in the first 
place (to lift the zero modes). 

Recall that under the shift symmetry, the 
$S^1$-component $B^{p-1}$ of $\mathscr{B}$ 
transforms as \eqref{deltaBC}
\be
\delta B^{p-1} = d_{A}^{N} \Sigma_{p-2} + (-1)^{p-1} D \sigma^{p-1}
\ee
We will now show that we can use just the second term 
($\Sigma_{p-2}$ is not required) to achieve 
$\delta B^{p-1} = - B^{p-1}$. 
Then for the transformed field one has
$B^{p-1} + \delta B^{p-1}=0$, as desired. This means that we have to solve
the equation
\be
(-1)^{p}B^{p-1} =  D \sigma^{p-1}.\label{BDS1}
\ee
We will construct the explicit solution for $\sigma^{p-1}$ below, 
but we can see immediately that a necessary (and sufficient) condition for
a solution to exist is that the operator $D$ is invertible, i.e.\ has
no zero modes (with the  
appropriate periodic boundary conditions). Recalling that 
the determinant of $iD$ is the Ray-Singer Torsion $\tau_{S^1}(\phi+\mu, \EE)$, 
we see that this gauge choice is available precisely when the massive
Ray-Singer Torsion is non-zero. 

In order to solve \eqref{BDS1}, the first step is to 
write the operator $D$ as 
\be
D = \del_\theta + \phi + \mu = k(\theta)^{-1} \del_\theta\; k(\theta) \label{def-G}
\ee
with
\be
k(\theta) = 
\mathrm{P}\exp{\left(\int_{0}^{\theta}(\phi + \mu)(s)
  ds\right)}, \quad k(2\pi) \equiv k = h g
\ee
the path ordered exponential of $\phi + \mu$, with $h$ and $g$ the 
previously introduced holonomies of $\mu$ and $\phi$ respectively 
\eqref{ghdef}. We shall also write
\be
\rho(k) = h\rho(g)\;\;.
\ee
Then one finds by straightforward integration that the solution is
\be
\sigma^{p-1} = \rho\left(k(\theta)
\right)^{-1}\left[\int_{0}^{\theta}\rho\left(k(s)\right)
  (-1)^{p}B^{p-1}(s)ds + \gamma  \right] \label{sigma-sol}
\ee
where the integration constant matrix $\gamma$ is determined by the 
requirement of periodicity of $\sigma^{p-1}$ on $S^1$ to be
\be
\gamma = 
(-1)^{p}\left(h\rho(g)-1\right)^{-1} \int_{0}^{2\pi}
\rho\left(k(s)\right)    B^{p-1}(s)ds \label{int-const-gamma}
\ee
Note that it is here that the requirement that $D$ be invertible enters, 
through the equivalent condition that 
$\Det_{E \otimes L}{(1-\rho(k)^{-1})} \neq 0$, 
which is, as anticipated above, precisely the condition that 
the massive Ray-Singer Torsion on the circle does not vanish,
\be
\Det_{E\otimes L}{(1-\rho(k)^{-1})} \neq 0 \quad\LRa\quad
\tau_{S^1}(\phi + \mu, \EE) \neq 0 \;\;.
\ee

In any case, as announced, we have now established that we can then 
achieve the gauge fixing conditions (\ref{gfixing}).

As mentioned above, 
we have not used $\Sigma_{p-2}\, d\theta$ thus far; 
however, by the second
symmetry (\ref{second-symm}) we can choose $\sigma^{p-2}$ to set
$\Sigma_{p-2}=0$ (as we now know that for such a shift symmetry this
gauge choice is achievable). If $p=2$ we have finished but if $p\geq
3$ then we have not made use of $\Sigma_{p-3}\, d\theta$ but fortunately we have
a next level of gauge symmetry and we can use the parameter of the next
level down with degree $(p-3)$ to set $\Sigma_{p-3}=0$ and so on all
the way until the gauge parameter has degree 0 - which means that there
is no $d\theta$ component and so nothing to gauge fix to zero.

We can summarise this by saying that we can use (and only need) the 
shift gauge symmetry parameters $\sigma^k$ 
with $k=0,\ldots,p-1$, 
to set to zero the $d\theta$-component $B^{p-1}$ of $\mathscr{B}^p$ 
and the $d\theta$-components $\Sigma_k$
of the shift gauge symmetry parameters themselves, 
\be
\label{completeagf}
\{\sigma^k, k=0,\ldots,p-1 \} \quad\Ra\quad B^{p-1}=0\quad,\quad \Sigma_k = 0
\quad k=0,\ldots,p-2\;\;.
\ee

\subsection{Path Integral Derivation of the Massive Ray-Singer Torsion}\label{PI-Massive-RST}

In this section we will explicitly evaluate the partition function
of the field theory models for the massive Ray-Singer Torsion
introduced above in the algebraic gauge. As a warm-up, and as it
is already quite instructive in its own right, we first discuss two
3-dimensional examples, in order to illustrate how
\begin{itemize}
\item the same ratio of determinants (of Laplacians) emerges from the 
usual covariant gauge (involving derivatives of the fields) 
and the algebraic gauge;
\item
in the algebraic gauge one can work with a significantly reduced field
content compared with the covariant gauge (which requires the full field
content of ghosts, anti-ghosts and ghosts for ghosts and their multipliers, 
as specified by the Batalin-Vilkovisky (BV) triangle \cite{BV1983}). 
\end{itemize}

\subsubsection{3-Dimensional Examples: Covariant vs Algebraic Gauge}

We consider the action \eqref{RST-action} on a 3-manifold of the
form $M = N \times S^1 \equiv \Sigma \times S^1$, 
with the form degree $(p,q)$ of
$(\mathscr{B},\mathscr{C})$ chosen to be $(p,q)=(1,1)$ and $(p,q)=(2,0)$
respectively.  The first case is irreducible and in a sense the
square of Abelian (or 1-loop) Chern-Simons theory \cite{Witten-CS}, while the second
case, with $\mathscr{B}$ a 2-form, is reducible.

\begin{example}{The classical action is 
\[
S_M(A) = \int_M \langle \mathscr{B}^1,d_A\mathscr{C}^1\rangle
\]
The shift gauge 
transformation $\delta \mathscr{B}^1=d_A\Sigma^0$ \eqref{RST-symm} 
translates into the BRST-symmetry $Q\mathscr{B}^1=d_A\Omega^0$ 
(and analogously for the 1-form $\mathscr{C}^1$). 
The complete BRST algebra for the fields $\mathscr{B}^1$ and $\mathscr{C}^1$
is then (regardless of the gauge fixing conditions)
      \begin{align}
   &   Q\mathscr{B}^{1} = d_{A} \Omega^{0}, \quad Q \Omega^{0} = 0 , \quad Q
      \overline{\Omega}^{0} = \Pi^{0}, \quad Q\Pi^{0}=0 , \quad Q^{2}=0\nonumber\\
   &  Q\mathscr{C}^{1} = d_{A} \Psi^{0}, \quad Q \Psi^{0} = 0 , \quad Q
      \overline{\Psi}^{0} = \Phi^{0}, \quad Q\Phi^{0}=0 , \quad Q^{2}=0\nonumber
      \end{align}
The usual covariant gauge fixing condition \eqref{cov-gf2} is imposed
by adding to the classical action 
$Q\int_M \langle \bar{\Omega}^0,d_A*\mathscr{B}^1\rangle 
+ Q\int_M \langle \bar{\Psi}^0,d_A*\mathscr{C}^1\rangle$, 
so that the complete quantum action is
\[
\int_M \left( 
\langle \mathscr{B}^1,d_A\mathscr{C}^1\rangle
+ \langle \Pi^0, d_A* \mathscr{B}^1\rangle +\langle  \bar{\Omega}^0, 
d_A*d_A\Omega^0\rangle
+ \langle \Phi^0, d_A* \mathscr{C}^1\rangle +\langle  \bar{\Psi}^0, 
d_A*d_A\Psi^0\rangle
\right)
\]
The integral over the Grassmann-odd ghosts and anti-ghosts
gives a $(\Det_{\Omega^0(M, \EE \otimes \mathbb{L})}\Delta_A)^2$, while the integral
over the Grassmann-even fields $\mathscr{B},\mathscr{C},\Pi^0,\Phi^0$
gives the determinant of the operator $L_A= *d_A + d_A*$ on odd forms
(say). By squaring this operator and taking the square root (unlike in
\cite{Witten-CS} there is no phase to worry about because of the 
doubled field content and on including the complex conjugate field set
if there is one), and using Hodge duality, 
one finally finds that the partition function is
precisely equal to the Analytic Torsion, 
\[
Z_M[A] = 
(\Det_{\Omega^0(M,\EE \otimes \mathbb{L})}\Delta_A)^{3/2} (\Det_{\Omega^1(M,\EE \otimes \mathbb{L})}\Delta_A)^{-1/2}
= \tau_M(A,\EE \otimes \mathbb{L})\;\;,
\]
in agreement with \eqref{ZRST}. 
On the product manifold $M= \Sigma\times S^1$, one could now use the product formula
\eqref{product-RST2} 
and the triviality of the Analytic Torsion on $\Sigma$ to deduce that 
\[
Z_{\Sigma \times S^1}[A] = \tau_{S^1}(\phi+\mu,\EE)^{e(\Sigma)}\;\;.
\]
We will now show how to obtain this formula directly by evaluating the path 
integral on $\Sigma\times S^1$ in the algebraic gauge $B^0=C^0=0$ \eqref{gfixing}, 
where $\mathscr{B}^1 = B^1 + B^0 d\theta$, $\mathscr{C}^1 = C^1 + C^0 d\theta$ 
(and we recall that for 0-form fields on $\Sigma\times S^1$ we have $\Omega^0=
\omega^0$ etc.). 
These algebraic conditions are thus imposed by adding to the classical action 
$Q\int_{M} * \langle B^{0}, \overline{\omega}^{0}\rangle 
+ Q\int_{M} * \langle C^{0}, \overline{\psi}^{0}\rangle$,  
leading to the complete action
\[
S_M(A) + \int_M * \left( 
\langle B^0,\pi^0\rangle + \langle D\omega^0,\overline{\omega}^0\rangle
+
\langle C^0,\phi^0\rangle + \langle D\psi^0,\overline{\psi}^0\rangle
\right)
\]
This clearly sets $B^{0}=C^0=0$ and integrating out the ghosts gives
      rise to the square of the 
determinant $\Det_{\Omega^{0}\left(S^1, \mathbb{E}
          \otimes \mathbb{L}\right) \otimes
        \Omega^{0}\left(\Sigma\right)}{(iD)}$. The classical action, on the 
other hand, reduces to $\int_M d\theta B^1 D C^1$ \eqref{RST-action-1}, 
and thus gives rise to  the inverse of the 
determinant $\Det_{\Omega^{0}\left(S^1, \mathbb{E}
          \otimes \mathbb{L}\right) \otimes
        \Omega^{1}\left(\Sigma\right)}{(iD)}$. 
Putting all the pieces together we get
      \[
Z_M[A]= \Det_{\Omega^{0}\left(S^{1}, \mathbb{E}\otimes
          \mathbb{L}\right)}{(iD)}^{2\dim{\Omega^{0}\left(\Sigma\right)
        } - \dim{\Omega^{1}\left(\Sigma\right)}} 
\]
At a heuristic level, one has (from the Hodge decomposition) 
\[
2\dim{\Omega^{0}\left(\Sigma\right)} - \dim{\Omega^{1}\left(\Sigma\right)}
= 2 \dim{ H^0(\Sigma)} - \dim {H^1(\Sigma)} = e(\Sigma)\;\;, 
\]
so that 
\[
Z_{\Sigma \times S^1}[A] = \Det_{\Omega^0\left(S^1, \EE \otimes \mathbb{L}\right)}
{(iD)}^{e(\Sigma)} = \tau_{S^{1}}(\phi +
      \mu, \mathbb{E})^{e(\Sigma)}
      \]
}
\end{example}

A correct account requires a regularisation of the objects involved
as given in \cite{BT-CS,BT-Trieste-1993}. We will come back to this
in Section \ref{subsubag} below. In any case, we have seen that
(modulo this proviso) in the algebraic gauge we have reproduced,
on the nose, the result obtained before from combining the calculation
in the covariant gauge with the product theorem for the Analytic
Torsion.

We now look at an example that illustrates the issues that arise
when the theory is reducible.

\begin{example}{Once more let $M= \Sigma \times S^{1}$ but with and $(p,q)= (2,0)$ in
      which case $\mathscr{B}$ is a 2-form while  $\mathscr{C}$ is a 0-form. 
The BRST algebra for the field $\mathscr{B}$ is
      \[
      Q\mathscr{B}^{2} = d_{A} \Omega^{1},  \quad Q \Omega^{1} = d_{A}\Omega^{0} , \quad Q
      \overline{\Omega}^{1} = \Pi^{1},  \quad Q
      \overline{\Omega}^{0} = \Pi^{0}, \quad  Q\gamma^{0} = \tau^{0}  \quad Q^{2}=0
      \]
      and which holds once more irrespective of the gauge fixing
      condition. There are no transformations for
      $\mathscr{C}^{0}$. Here $\Omega^0$ is the ghost for ghost, 
required due to the reducibility of the transformation 
$Q\mathscr{B}^2 = d_A \Omega^1$, and $\overline{\Omega}^0$ is its anti-ghost.  
The fields $(\tau^0,\gamma^0)$ are the extra ghost and its multiplier, 
required in general in the presence of this reducible symmetry, with $\tau^0$
completing the BV triangle:
\[
\begin{array}{ccccc}
&&\mathscr{B}^2&&\\
&\overline{\Omega}^1 & & \Omega^1 & \\
\tau^0 && \overline{\Omega}^0 && \Omega^0
\end{array}
\]
We will see now that (and why) $\tau^0$ is indeed required in the
covariant gauge, while it is not required in the algebraic gauge. 
This generalises to the higher-dimensional theories and higher rank 
forms and will significantly simplify the calculation of the 
partition function in Section \ref{subsubag} below.

In the covariant gauge, in order to gauge fix the shift gauge symmetry
of the field $\mathscr{B}^2$ and its ghost $\Omega^1$, one can add to the 
action the terms
$Q\int_M \langle \bar{\Omega}^1,d_A*\mathscr{B}^2\rangle + 
Q\int_M \langle \bar{\Omega}^0,d_A*\Omega^1\rangle$. However, one 
now sees that the multiplier $\Pi^1$, which only appears in 
the combination $\int_M \Pi^1 d_A*\mathscr{B}^2$, has its own 
gauge invariance (as the $d_A$-exact part of $\Pi^1$ does not appear
in the action). This is the rationale for the necessity of the 
extra ghost $\tau^0$ and its multiplier $\gamma^0$ in the covariant 
gauge, as these allow one 
to gauge fix $\Pi^1$ (and its multiplier $\overline{\Omega}^1$), by 
adding a third covariant gauge fixing term, namely
\[
Q \int_M \langle \gamma^0, d_A*\overline{\Omega}^1\rangle = 
\int_M \langle \tau^0, d_A*\overline{\Omega}^1\rangle + 
\int_M \langle \gamma^0, d_A*\Pi^1\rangle\;\;.
\]
The resulting partition function is now well defined and 
can be evaluated much like in the previous example, with 
the result 
\[
Z_M[A] = 
(\Det_{\Omega^0(M,\EE \otimes \mathbb{L})}\Delta_A)^{-3/2} 
(\Det_{\Omega^1(M,\EE \otimes \mathbb{L})}\Delta_A)^{+1/2}
= \tau_M(A,\EE \otimes \mathbb{L})^{-1}\;\;,
\]
once again in agreement with \eqref{ZRST}. Using again the product formula, 
on $M = \Sigma \times S^1$ this is equal to 
\[
Z_{\Sigma \times S^1}[A] = \tau_{S^1}(\phi+\mu,\EE)^{-e(\Sigma)}\;\;.
\]

Working now directly on $\Sigma \times S^1$ instead, recall that we have the 
decompositions
$\mathscr{B}^2=B^2 + B^1d\theta$, $\Omega^1 = \omega^1 + \Omega_0 d\theta$, 
$\Pi^1 = \pi^1 + \Pi_0 d\theta$ etc., as well as $\Omega^0 = \omega^0$ etc.\ 
for 0-form fields.  
The algebraic gauge conditions on the 
fields $\mathscr{B}^2$ and its ghost $\Omega^1$ are the conditions
$B^1=0, \Omega_0=0$, which can be imposed by adding 
$Q\int_{M} (\langle B^{1}, * \overline{\Omega}^{1}\rangle 
+ \langle \Omega_{0}, * \overline{\Omega}^{0} \rangle)$
to the action. However, because $B^1$ is a 1-form on $\Sigma$, 
$\langle B^1, *\overline{\Omega}^1\rangle = 
\langle B^1, *\overline{\omega}^1\rangle$, and therefore 
at this stage neither $\overline{\omega}_0$ nor its multiplier $\pi_0$ will 
appear anywhere in the action. If one insists on using
the full BV field content, then one can once again use the 
ghost for ghost $\tau^0$ and its multiplier to set these fields to 
zero (algebraically), by adding $Q\int_M \gamma^0*\overline{\Omega}_0$. Using 
also that 
\[
Q\mathscr{B}^2 = d_A \Omega^1 \quad\Ra\quad 
QB^1 = d_A^\Sigma\Omega_0 + D\omega^1\;\;,
\]
one then finds that the terms that need to be added to the classical
action to arrive at a well-defined partition function are 
      \begin{align*}
      &  \int_{M} Q\left(\langle B^{1}, *
        \overline{\omega}^{1}\rangle + \langle \gamma^{0}, *
        \overline{\Omega}_{0}\rangle + \langle \Omega_{0}, *
        \overline{\omega}^{0} \rangle\right)\\
       & = \int_{M} \left( \langle B^{1}, *
        \pi^{1}\rangle + \langle (d^{\Sigma}_{A}\Omega_{0} + D
        \omega^{1}), *
        \overline{\omega}^{1}\rangle + \langle \tau^{0}, *
        \overline{\Omega}_{0}\rangle 
           + \langle \gamma^{0}, *
        \Pi_{0}\rangle + \langle \Omega_{0}, *
        \pi^{0} \rangle + \langle D\omega^{0}, *
        \overline{\omega}^{0} \rangle\right)
      \end{align*}
After integrating out all the multiplier fields imposing the algebraic 
gauge conditions, one is then (including the classical action) 
left with 
\[
\int_M d\theta \langle B^2,DC^0\rangle 
+ \int_M \langle D\omega^1, *\overline{\omega}^1\rangle
+ \int_M \langle D\omega^0, *\overline{\omega}^0\rangle
\]
This immediately leads to 
\[
Z_{\Sigma \times S^1}[A] = 
      \Det_{\Omega^{0}\left(S^{1}, \mathbb{E}\otimes
          \mathbb{L}\right)}{(D)}^{-\dim{\Omega^{2}\left(\Sigma\right)
        } + \dim{\Omega^{1}\left(\Sigma\right)} -
        \dim{\Omega^{0}\left(\Sigma\right)}} = \tau_{S^{1}}(\phi + 
      \mu, \mathbb{E})^{-e(\Sigma)} 
      \]
(using the same heuristic reasoning as before), in agreement with the 
result found in the covariant gauge. 
    }
\end{example}

While this is somewhat satisfactory, it is also clear from the
derivation that the procedure we have followed here in the second example
(with its full
field content as dictated by the BV triangle) is overly baroque and
complicated. Indeed, if only the field $\pi^1$ in the decomposition
$\Pi^1 = \pi^1 + \pi_0 d\theta$ of the multiplier field is required
to impose the algebraic gauge condition $B^1=0$, then there is no
reason to introduce $\pi_0$ in the first place. And if one does not
introduce $\pi_0$, then one does also not need a multiplier $\gamma^0$
(and its extra ghost) to set it (and its anti-ghost) to zero. The upshot
is that if one simply does not introduce those fields, one does not need
the extra ghost.

\subsubsection{General Calculation in the Algebraic Gauge}
\label{subsubag}

We will now show that this kind of reasoning can be extended to
arbitrary $(p,q=n-p-1)$, where it leads to a rather more significant
simplification of the calculation of the partition function than
in the above (after all barely non-trivial) 3-dimensional example.

For reducible gauge systems (displaying the ghost for ghost and extra
ghost phenomena), the Batalin-Vilkovisky triangle \cite{BV1983} has
a right edge which includes the classical field to be quantised as
well as the ghost terms. All the other fields in the triangle are
the anti-ghosts and extra ghosts. Each of the anti-ghosts and extra
ghosts have matching multiplier fields.
In the case at hand, the $\mathscr{B}^p-\mathscr{C}^q$ system, we
would have BV triangles of height $p+1$ and $q+1$ respectively,
with a corresponding explosion in the number of fields. 

However, for an algebraic gauge condition the only place from which
a non-trivial determinant (involving derivatives of the fields) can
arise is from the classical action and from the terms involving the
BRST variations of the classical fields and the ghosts  on the right
edge of the triangle. These only couple to the anti-ghosts and their
multiplier fields. Thus only the fields that appear on the first
and second right edges of the triangle (and their multipliers)
are required, as indicated here for the field $\mathscr{B}^p$:
\[
\begin{array}{cccccc}
&&\mathscr{B}^p&&&\\
&(\overline{\Omega}^{p-1},\Pi^{p-1}) & & \Omega^{p-1} && \\
&& (\overline{\Omega}^{p-2},\Pi^{p-2}) && \Omega^{p-2}&\\
&&& \cdots && \cdots
\end{array}
\]

Concretely, since we just need the $\pi^i$-components of the multiplier
fields $\Pi^i=\pi^i + \Pi_{i-1}$ to impose the algebraic gauge conditions
$B^{p-1}=0, \Omega_{i}=0$, the reduced set of fields is the minimal one 
required for the gauge fixing as indicated in \eqref{completeagf}. 
The 
BRST transformations on this reduced set of 
fields in the $\mathscr{B}^p$-sector are simply
\be
Q\Omega_{i} = D\omega^{i}, \quad Q\omega^{i}=0, \quad
Q\overline{\omega}^{i} = \pi^{i}, \quad Q\pi^{i}=0, \quad Q^{2}=0
\ee
with $\Omega_{p-1}=B^{p-1}$ and $i = 0, \dots , p-1$. The simplified
set of gauge fixing conditions are 
\be
 \sum_{i=0}^{p-1} \int_{M} Q\langle \Omega_{i}, * \overline{\omega}^{i}
 \rangle = \sum_{i=0}^{p-1} \int_{M} \left(\langle \Omega_{i}, * \pi^{i}
 \rangle + \langle D\omega^{i}, * \overline{\omega}^{i}
 \rangle\right) \label{gen-algebraic}
\ee
From the various fields we get the following determinants:
\begin{equation}
\begin{aligned}
(B^p,C^{n-p-1}) \quad & \Ra \quad
\Det_{\Omega^{0}\left(S^{1}, \mathbb{E}\otimes
          \mathbb{L}\right)}{(D)}^{-\dim{\Omega^{p}\left(N\right)}}  \\
\mathscr{B}\; \mathrm{ghosts} \quad & \Ra \quad
\Det_{\Omega^{0}\left(S^{1}, \mathbb{E}\otimes
          \mathbb{L}\right)}{(D)}^{\sum_{i=0}^{p-1}
        (-1)^{p+1+i}\dim{\Omega^{i}\left(N \right)
        }} \\ 
\mathscr{C}\; \mathrm{ghosts} \quad & \Ra \quad
      \Det_{\Omega^{0}\left(S^{1}, \mathbb{E}\otimes
          \mathbb{L}\right)}{(D)}^{\sum_{i=0}^{n-p-2}
        (-1)^{n-p+i}\dim{\Omega^{i}\left(N \right)
        }} 
\end{aligned}
\end{equation}
(we have assumed that both $\mathscr{B}$ and $\mathscr{C}$ are
Grassmann even -  otherwise one would have to simply 
invert this result). One can
use duality to set $\sum_{i=0}^{n-p-2}
(-1)^{n-p+i}\dim{\Omega^{i}\left(N \right)} = \sum_{i=p+1}^{n-1}
(-1)^{p+1+i}\dim{\Omega^{i}\left(N \right)}$, so that the product of
the determinants is the partition function
\be
Z_{\Sigma \times S^1}[A] = \Det_{\Omega^{0}\left(S^{1}, \mathbb{E}\otimes
          \mathbb{L}\right)}{(D)}^{(-1)^{p+1} \sum_{i=0}^{n-1} (-1)^{i}
        \dim{\Omega^{i}\left(N\right) 
         }}
       \ee
As before, formally the exponent is (up to a sign) the 
Euler characteristic $e(N)$ 
of $N$. 

We now explain how meaning is given to these
formulae. Clearly one must regularise the fact that one is dealing
with all forms on $N$. The McKean-Singer formula \cite{McKeanSinger} uses a Heat Kernel
regularisation for these objects. Namely one writes \cite{BT-CS}
\[
\ln{\left(\tau_{S^{1}}(\phi+\mu, \mathbb{E})^{\sum_{i=0}^{n-1} (-1)^{i}
        \dim{\Omega^{i}\left(N\right) 
        }}\right)} = \lim_{\eps \rightarrow 0}
  \sum_{i}\Tr_{\Omega^{i}(N)}{(-1)^{i}\ln{\tau_{S^{1}}(\phi+\mu, \mathbb{E})}
    \ex{-\eps \Delta_{A}}}
  \]
  We use the obvious (position space) notation for the functional trace, 
  \be
   \Tr{\ln{\tau_{S^{1}}(\phi+\mu,
     \mathbb{E})} 
   \ex{-\eps \Delta_{A}}} 
 = \int_{N}
  \ln{\tau_{S^{1}}(\phi+\mu, \mathbb{E})(x)} <x|\ex{-\eps \Delta_A}
  |x>
\ee
The coefficients 
arising
in the expansion of the heat kernel are known to be local and gauge invariant, and since
$F_A=0$ we can replace $\Delta_A \ra \Delta$ in the above calculation. We
can then make use of the fact that 
the index density of the deRham operator is the Pfaffian of the Riemann 
curvature tensor (see e.g.\ equation (1) in \cite{Patodi-1971}), 
\be
\lim_{\eps \rightarrow
   0}\sum_{i}(-1)^{i}<x|\ex{-\eps \Delta}
 |x> = \mathrm{Pf}(R)
 \ee
Thinking of $\mathrm{Pf}(R)$ as the Pfaffian of the 
$\mathfrak{so}(n-1)$-valued Riemann curvature 2-form of $N$ 
(thought of as a top-form on $N$), we can write
  \be
  \lim_{\eps \rightarrow
   0}\sum_i \Tr_{\Omega^{i}(N)}{(-1)^{i}\ln{\tau_{S^{1}}(\phi+\mu,
     \mathbb{E})} 
   \ex{-\eps \Delta_{A}}} 
  = \int_{N}
  \ln{\tau_{S^{1}}(\phi+\mu, \mathbb{E})} \mathrm{Pf}(R)
  \ee
The partition function is, therefore,
  \be
  Z_{\Sigma\times S^1}[A] = \exp{\left((-1)^{p+1}\int_{N}\ln{\tau_{S^{1}}(\phi+\mu, \mathbb{E})}
    \mathrm{Pf}(R)\right)}
  \ee
  In case that $\phi+\mu$ are constant on $N$ we have
  \be
  Z_{\Sigma \times S^1}[A] = \tau_{S^{1}}(\phi+\mu, \mathbb{E})^{(-1)^{p+1}e(N)}
  \ee
  justifying our rather cavalier formulae in the examples above.    

\subsection{Remarks and Observations}
      
\subsubsection{Independence of the  Connections on $N$}

Notice that the result for the Analytic Torsion does not depend
explicitly on the component of the flat connection $A_{N}$ on $N$. This rather surprising
result also is a generalisation of the results of Ray and
Singer. In order to see that it is not obvious that the component of
the connection on $N$ does not appear we take a quick look at
the flat connections on $N \times I$ and those that are actually flat
on $N \times S^{1}$ to show that there is no apriori reason to ignore $A_{N}$.

The flat connections on $N \times I$, with the interval $I = [0, 2\pi]$, satisfy
\be
F_{A}^{N}=0, \quad \dot{A}_{N} = d_{A}^{N}\phi
\ee
where the superscript $N$ indicates that the exterior derivative and
connection are those on $N$. The first equation says that $A_{N}$ must
be a flat connection on $N$ for all `time' $\theta$ while the second says
that the time evolution of $A_{N}$ is given by a gauge transformation
parameterised by $\phi$. The evolution equation is solved by
\be
A_{N}(x,\theta) = g^{-1}(x,\theta ) A_{N}(x,0) g(x, \theta) +
g^{-1}(x,\theta)d^{N}g(x,\theta) \label{flat-NI}
\ee
where
\be
g(x,\theta) = \mathrm{P} \exp{\left(\int_{0}^{\theta}\phi(x,s) ds\right)}
\ee
and to satisfy the first equation $A_{N}(x,0)$ is a flat connection on
$N$.

One passes to flat connections on $N \times S^{1}$ by imposing
the periodicity conditions
\be
A_{N}(x, 2\pi) = A_{N}(x, 0), \quad \phi(x, 2\pi) =
\phi(x, 0)
\ee
This gives us, from (\ref{flat-NI}), that $A_{N}(x,0)$ satisfies
\be
A_{N}(x,0) = g^{-1}(x, 2\pi ) A_{N}(x,0) g(x, 2\pi) +
g^{-1}(x, 2\pi)d^{N}g(x,2\pi) 
\ee
We have two extremes, the first is with $\phi=0$, for which this
equation gives no condition and $A_{N}(x,0)$ any flat connection on
$N$ while the second is with $A_{N}(x,0)=0$ and then $\phi$ is
constant on $N$ and hence a flat connection on 
$S^{1}$. For inbetween values we have that $A_{N}(x,0)$ is reducible
and the reducibility is determined by the holonomy on $S^{1}$ of $\phi$.

Consequently we see that there are, in principle, many connections
that require both $A_{N}$ and $\phi$ to be non-zero and so it is a
non-empty result that the torsion does not depend on $A_{N}$.

\subsubsection{Application to Abelianisation}
\label{subsubabel}

Quite surprisingly the fact that $A_{N}$ does not make an
appearance in the formulae for the Ray-Singer Torsion also puts
into context a perplexing aspect of the Abelianisation programme
initiated in \cite{BT-CS,BT-Trieste-1993,BT-DIA}, and applied to 
the evaluation of the Chern-Simons partition function on various
classes of 3-manifolds of increasing complexity 
in \cite{BT-CS, BT-S1, BT-Seifert, BKNT}.

Concretely, in Chern-Simons theory (or its relatives such as $BF$
theory and the $G/G$ gauged sigma model) one is able to choose a
gauge condition that Abelianises certain fields so that, within the
evaluation of the associated path integral, one encounters a
Ray-Singer Torsion that is a function only of that Abelianised field,
even though that Abelian field is typically, by itself, not a flat
connection.
However, here we see that $\phi$ (and $\mu$) while they may not be
flat by themselves, can be regarded as components of a flat connection
on $M$ which also includes a non-trivial $A_{N}$, while the result
does not depend on the explicit form of $A_N$ at all. 

In order to fix ideas, consider the Chern-Simons path
integral on $M=\Sigma \times S^{1}$ with $\Sigma$ a compact closed
Riemann surface \cite{BT-CS}. Let $\phi$ be the component of
the connection along the $S^{1}$ direction. After integrating over the
other components of the connection as well as the ghost terms one
finds that the absolute value of the partition function is
proportional to the Ray-Singer Torsion (there is still an integral
over $\phi$ to perform)
\be
\tau_{S^{1}}( \phi, \lk)^{e(\Sigma)/2}\;\;.
\ee
Since we have already integrated over all the components of the connection
$A_{\Sigma}$,  it is not at all obvious that $\phi$ is a component of some
flat connection on $M$. However, as we have seen, $\phi$ can be
thought of as a component of a flat connection
$(\widetilde{A}_{\Sigma}, \phi)$ as the Ray-Singer Torsion will not, in
any case, depend on the rest of the flat connection. This in turn
means that one may therefore compare with the traditional approach of
determining the Chern-Simons path integral which involves an expansion
about the flat connections. The crucial advantage here is that we do not 
actually need 
to know the flat connections completely, 
but only their possible $\phi$ components.

\subsubsection{Metric Independence} 
\label{subsubmiya}

The pairing that one uses for the ghost terms in the gauge fixing
procedure makes use of the Hodge star, so of the metric on the
underlying Riemannian manifold. One need not do this. For example one
could in QED consider the ghost $\omega$ to be a 0-form while taking
both the
anti-ghost $\overline{\omega}$ and multiplier field $\pi$ to be
$n$-forms. In this way the covariant gauge fixing and ghost terms would be
\be
\int_{M} \left(\langle \delta A, \pi\rangle + \langle
        \omega, \Delta \overline{\omega} \rangle  \right)
\ee
Usually, nothing is gained by doing this as the Hodge star resides in
the Laplacian as well.

However, with the algebraic conditions (\ref{gen-algebraic}) if one opts
to use the Hodge duals of the anti-ghosts and multiplier fields then
one sees quite straightforwardly that there is no metric at all in the
gauge fixed action. But what of the path integral measure? We note
that the anti-ghosts and the multiplier fields are always evenly
matched and always of opposite statistics so that they form a type of
supersymmetric pair. Consequently, the easiest way to see the metric
independence of the theory is to use the fields adapted so there is no
explicit metric dependence in the action.

We should also make a comparison with the path integrals of Section
\ref{MRSTS1}. So consider $N$ to be a point and suppose that $p=0$.
Then the action that we should consider to arrive at
the torsion is the Grassmann odd version of the action \eqref{RST-action-1}
with $(B,C) \rightarrow (\overline{\eta}, \eta)$, i.e.
\be
(-1)^{p}\int_{M}d\theta \langle B^{p},
  DC^{q} \rangle  \quad \ra\quad 
\int_{S^{1}} d\theta \langle \overline{\eta}, D \eta \rangle\;\;,
\ee
with $\overline{\eta}$ and $\eta$ being Grassmann zero-forms on
$S^{1}$. There is no gauge fixing
involved and so the theory is metric independent from the start and
coincides with that considered in Section \ref{MRSTS1} (after we set $R=1$).

\subsubsection{From 2nd to 1st Order Actions on $N\times S^1 $}

We will give yet another derivation of the formulae for the torsion on
$N \times S^{1}$ with metric $g_{M}= g_{N} + R^{2} d\theta \otimes
d\theta$ and a background $\mathbb{R}_{+}$ connection 
switched on. The standard Schwarz path integral for the Ray-Singer
Torsion requires an
interpretation of the determinants of first order operators $T$ as
being formally given as $\Det{(TT^{\dagger})}$ with the adjoint
operator $T^{\dagger}$ being a metric adjoint. Our aim here is to show
that the ratio of determinants of Laplacians, that goes into the
definition of the Ray-Singer Torsion, devolves to the ratios of the
determinants of the first order operators of the previous sections,
completely by-passing the need to give a separate definition for the
determinants.

This derivation requires that the overall path integral
is metric independent and that we may take the singular limit $R
\rightarrow 0$. The simplest way to ensure this is to consider the
massive theory as there are no zero modes to contend with. We will make use of the knowledge that we have gained
previously in that rather than the mass we should use $\mu$ and on the
circle there is no metric dependence on using this variable, in
particular it is $R$ independent.

We begin with the square of the product of determinants that go into
the definition (\ref{mass-RST}) with mass 
\be
\widehat{\tau}^{2}_{N\times S^{1}}(A,m)= \prod_{j=0}^{n} 
\left(\Det_{\Omega^{j}(M,\mathbb{E})}{(\Delta_{A}+ m^{2}
    )}\right)^{(-1)^{j+1}j}\label{massive-S1}
\ee
Now use a standard representation of the determinant of the
Laplacian as a Gaussian field theory so that we may express
(\ref{massive-S1}) in terms of path integrals as
\be
\widehat{\tau}^{2}_{M}(A,m) = \prod_{i} \int DB DC \,
\exp{\left(i\int_{M} \langle B,  (\Delta_{A} + m^{2}) C\rangle \right) } \label{Z-KK}
\ee
where the product is over $i$ copies of each set of $(i, n-i)$ forms $(B^{i},
C^{n-i})$ and the pairs are Grassmann even if $B^{i}$ is of even
degree while they are Grassmann odd if $B^{i}$ is of odd degree. This
product, representing the Ray-Singer Torsion, will be metric independent.

For every Gaussian path integral factor on $M=N \times S^{1}$ we split the Laplacian as
follows
\[
\exp{\left(i\int_{M} \langle B,
  \left( \Delta_{A}^{M} + m^{2}\right) C\rangle \right) } =
\exp{\left(i\int_{M} \langle B, 
  \Delta_{A}^{N} C\rangle \right) }\exp{\left(i\int_{M} \langle B,
  (\Delta_{A}^{S^{1}}+m^{2}) C\rangle \right) }
\]
and introduce a second path integral
\begin{align}
 & 
\exp{\left(i\int_{M} \langle B,
    (\Delta_{A}^{S^{1}}+m^{2}) C\rangle \right) }\nonumber \\
 & 
\quad \quad = \int D\widehat{B} D \widehat{C}
\, \exp{\left( i\int_{M} \left( \langle B,  D_{\phi+\mu} \widehat{C}\rangle
 +  \langle \widehat{B},  D_{\phi-\mu} C\rangle + R^{2}\langle \widehat{B},
 \widehat{C} \rangle \right)\right)} 
\end{align}
Notice that in this formula we have once more set $\mu = Rm$. We
substitute back into (\ref{Z-KK}) to arrive at 
\begin{align}
\widehat{\tau}^{2}_{M}(\phi, \mu) &=  \prod_{i} \int DB DC
D\widehat{B} D \widehat{C} \nonumber\\
& \quad 
\exp{\left(i\int_{M} \langle B,  \Delta_{A}^{N} C\rangle + \langle B,
    D_{\phi+\mu} \widehat{C}\rangle 
 +  \langle \widehat{B},  D_{\phi-\mu} C\rangle + R^{2}\langle \widehat{B},
 \widehat{C} \rangle \right) } 
\end{align}
Now we take the $R \rightarrow 0$ limit while keeping $\mu$ fixed. As $D_{\phi \pm \mu}$ are
invertible we may ignore the term involving the Laplacian on $N$
altogether as the integrals over $(\widehat{B}, \widehat{C})$
essentially set $(B,C)$ to zero. We thus arrive at
\be
\widehat{\tau}^{2}_{M}(\phi, \mu) = \prod_{i}\int DB DC
D\widehat{B} D \widehat{C}\exp{\left(i\int_{M}  \langle B,
    D_{\phi+\mu} \widehat{C}\rangle 
 +  \langle \widehat{B},  D_{\phi-\mu} C\rangle \right) } 
\ee
We have therefore derived the representation of the Ray-Singer Torsion in terms of path integrals of 1st order
actions from its definition in terms of products of determinants of Laplace operators
(path integrals of 2nd order actions), thus in a sense reversing the usual derivation of the 
Ray-Singer Torsion from the 1st order Schwarz actions.

As an aside we note that it might seem strange that we are taking the
$R\rightarrow 0$ limit while keeping $\mu$ fixed which amounts to
simultaneously taking $m \rightarrow \infty$, while it may be the
intention to eventually take the $m \rightarrow 0$ limit. However, we
saw in Section \ref{MDA} that the limit is correctly reformulated in
terms of $\mu$ as then no metric dependence arises.

To complete the derivation we note that the forms in question are on
$M$ and that we should break them down to their form degree on $N$,
\be
\Omega^{i}(M, \mathbb{E}) = \Omega^{i}(N, \mathbb{E})\otimes
\Omega^{0}(S^{1}) \oplus \Omega^{i-1}(N, \mathbb{E})\otimes
\Omega^{1}(S^{1})
\ee
Under this decomposition we have
\be
(B^{i}, \widehat{C}^{n-i}) \rightarrow (D^{i}, \overline{E}^{n-i-1})
\oplus (\overline{D}^{i-1}, E^{n-i})
\ee
(at the same time leading to an explicit $d\theta$ in the action) and
a similar decomposition for the $(\widehat{B}^{i}, C^{n-i}) $ pair.

The Grassmann count for $(D^{i}, \overline{E}^{n-i-1})$ is $(-1)^{i}$
while that for the pair
$(\overline{D}^{i}, E^{n-i-1})$ the Grassmann degree is
$(-1)^{i+1}$. From the multiplicity of determinants in
(\ref{massive-S1}) the determinants of $D_{\phi+\mu}$ acting on $i$ forms are
$i$ of the type that arise with $(D^{i},
\overline{E}^{n-i-1})$ and $i+1$ of the type $(\overline{D}^{i},
E^{n-i-1})$ and these two types have opposite
statistics. Happily enough all the path integrals now have the form of the
previous section leading to an overall determinant
\begin{align}
\widehat{\tau}^{2}_{N\times S^{1}}(\phi, \mu)&= \prod_{i=0}^{n-1}\Det_{\mathbb{E} \otimes
  \Omega^{i}(N)}{\left(D_{\phi+\mu}  \right)}^{(-1)^{i}} \, . \, \Det_{\mathbb{E} \otimes
  \Omega^{i}(N)}{\left(D_{\phi -\mu}  \right)}^{(-1)^{i}}\nonumber\\
&= \widehat{\tau}_{N\times S^{1}}(\phi+\mu)\, .\,
\widehat{\tau}_{N\times S^{1}}(\phi-\mu) 
\end{align}
This formula generalises that derived when $N$ is a point (\ref{RST-fact}).

For $\phi \pm \mu$ constant and following the regularisation of previous sections
we arrive at
\be
\widehat{\tau}^{2}_{N\times S^{1}}(A,m) =
\widehat{\tau}_{S^{1}}(\phi + \mu,
\mathbb{E})^{e(N)}. \widehat{\tau}_{S^{1}}(\phi - \mu,
\mathbb{E})^{e(N)}
\ee

We end this section with one last comment on metric dependence. Had we
used the `physics' path integral measure then its conventional normalisation
would have been
\be
D(BR^{-1}) D(CR^{-1})
\ee
However, the correct measure for the $(\widehat{B}, \widehat{C})$
systems is
\be
D(\widehat{B}R) D(\widehat{C}R)
\ee
and the product measure is
\be
D(BR^{-1}) DC(R^{-1})D(\widehat{B}R) D(\widehat{C}R) = DB DC
D\widehat{B}D\widehat{C}
\ee
which is $R$ independent. The complete $R$ dependence is then in the
action and so taking the $R\rightarrow 0$ limit amounts to just
dropping the $R^{2}\langle \widehat{B}, \widehat{C} \rangle$ term.

\section{Massive Ray-Singer Torsion on Mapping Tori}\label{sec-map}

We now apply the techniques introduced previously to the case when
$M$ is a finite order mapping torus. The approach that we use is to
twist the fields by the finite order diffeomorphism along the
circle. We will simultaneously perform a change of variables that pushes
the mass dependence out of the action and into the `boundary
conditions' along the $S^{1}$, as we did in previous sections. In particular we will establish
a generalisation of Fried's formula \eqref{fried-map} \cite{Fried}.

At a technical level we will need to establish that the equivalent of
the temporal gauge still applies on these manifolds. We will also need to use a
Lefshetz version of the index theorem. 

\subsection{Fields on Mapping Tori}

Let $N$ be a smooth (compact, closed) oriented manifold 
and let $f\in \mathrm{Diff}(N)$ be a diffeomorphism of $N$. 
Then the \textit{mapping torus} $M=N_f$ of $f$ can be concretely
described as the manifold
\be
M = N_f \equiv \frac{N \times I}{(x,0) \sim (f(x),1)}
\ee
where $I=[0,1]$ is the unit interval. This shows that 
$N_f$ is a fibration $N\ra N_f \stackrel{\pi}{\ra} S^1$ over $S^1$, 
with $\pi(x,t) = t$ and typical fibre $N$. 

Fields $\Phi(x,t)$ on $N_f$ can be regarded 
\begin{itemize}
\item either as functions on $N\times I$ satisfying the 
twisted boundary condition 
\be
\Phi(x,0) = f^*\Phi(x,1) \label{period-NxI}
\ee
\item or as functions on $N \times \RR$ satisfying the twisted 
periodicity conditions
\be
\Phi(x,t) = f^*\Phi(x,t+1)
\ee
\end{itemize}
Likewise, connections and gauge transformations can be chosen to 
satisfy the same periodicity
rules
\be
A(x,t) =f^{*} A(x,t+1), \quad g(x,t) =f^{*} g(x,t+1) \label{periodic-gauge}
\ee

In the following, we will consider the case that $f$ has finite order
$n_{f}$, i.e.\ $f^{n_{f}} = \mathrm{Id}$ with $\mathrm{Id}$ the identity
map (note that $f^{n_{f}} = f \circ\ldots \circ f$ ($n_{f}$ times) refers to
the group structure in $\mathrm{Diff}(N)$). 
In this case, $N\times S^1$ (with the circle coordinate $t\in [0,n_{f})$)
is an $n_{f}$-fold cover of $N_f$. This is reflected in the fact that 
the $n_{f}$-fold iteration of the twisted periodicity conditions gives
\be
\Phi(x,t) = (f^*)^{n_{f}}\Phi(x,t+n_{f}) = \Phi(x,t+n_{f})
\ee
(and likewise for $A$ and $g$). 

As on $N\times S^1$, 
one may unambiguously decompose the connection as
\be
A(x,t) = A_{N}(x,t) + \phi\, dt \label{conn-decomp}
\ee
with $\iota_{\xi}A_{N}(x,t)=0$ with $\xi$ the vector field
$\partial/\partial t$.

\subsection{The Partition Function and Gauge Fixing}

The partition function of a theory on the mapping torus, as will be
considered by us below, is over the
class of fields on $N \times I$ that satisfy (\ref{period-NxI}). We
will write the partition function of the theory yielding the Analytic
Torsion on $M$ as
\be
Z_{N_f}= \int_{\Phi(0)=f^{*}\Phi(1)} D\Phi\, \exp{\left(S_{N \times I}(\Phi)\right)}
\ee
for
\be
S_{N \times I}(\Phi) = \int_{N\times I} \langle \mathscr{B}, d_{A} \mathscr{C}\rangle
\ee
The shift gauge symmetry
\be
\delta \mathscr{B}= d_{A}\Sigma, \quad \delta \mathscr{C}= d_{A}\Lambda
\ee
now requires that the symmetry parameters, as every other field, obey
the twisted boundary conditions
\be
\Sigma(0,x) =f^{*}\Sigma(x,1), \quad \Lambda(0,x) =f^{*}\Lambda(x,1)
\ee
We must check to see if the algebraic gauge fixing conditions that
were so effective on the product $N \times S^{1}$ can also be used on
the mapping torus $N \rightarrow M \rightarrow S^{1}$. This amounts to
showing that a solution to (\ref{BDS1}) exists within the set of forms with
twisted boundary conditions. The solution, without boundary conditions
on $N \times I$ remains (\ref{sigma-sol}) and our task is determine
the analogue of the integration constant (\ref{int-const-gamma}) for
the twisted boundary conditions. As the general case is a bit messy, we
only provide the solution for $\phi + \mu$ which are constant on $N$,  
and do so to convince the reader that this is indeed possible. It is
\be
\gamma = 
(-1)^{p}\left(h^{n_{f}}\rho(g)^{n_{f}}-1\right)^{-1} \sum_{i=0}^{n_{f}-1}
\left(h\rho(g)f^{*}\right)^{i}\int_{0}^{1}
h^{s}\rho\left(g\right)^{s}  B^{p-1}(s)ds \label{int-const-gamma-2}
\ee

Now that we know that we have the ability to gauge fix as before we
can go to the BRST structure. The ghost for ghost system that one has
in the mapping torus case is 
the same as that for the product manifold. In any case this means that we have
exactly the same system as before (\ref{gen-algebraic}) with the only
difference being that the 
fields $(\omega^{i}, \overline{\omega}^{i}, \pi^{i})$ now also satisfy
the twisted boundary conditions. 

\subsection{Path Integral Derivation of  a Generalisation of Fried's Formula}\label{pidgff}

After gauge fixing we are confronted with products of `independent'
partition functions of the form
\be
Z_{i} =\int D\alpha D\beta
\; \ex{\left(i \int_{N \times I} dt
  \langle \alpha(t) ,\, D \beta(t) \rangle \right)}\, \delta\left(
\alpha(0)-f^{*}\alpha(1) \right)\, \delta\left(
\beta(0)-f^{*}\beta(1) \right)
\ee
with $\alpha \in \Omega^i(N, \mathbb{E})\otimes \Omega^0(I) $,
the appropriate boundary  
conditions on the mapping torus being specified and implemented by the 
explicit $\delta$-function insertions in the path integral.
By using (\ref{def-G}), one can now write the action as
\be
\int_{N \times I} 
  \langle \alpha(t) ,\, k^{-1}(t)\frac{\partial}{\partial t} k(t)\beta(t)
  \rangle
  \ee
Changing variables to $\widetilde{\alpha}(t)= \alpha(t) k^{-1}(t)$ and
$\widetilde{\beta}(t)= k(t) \beta(t)$, the action simplifies to 
\be
 \int_{N \times I} 
  \langle \widetilde{\alpha} , \frac{\partial}{\partial t}
  \widetilde{\beta} \rangle
  \ee
  while the boundary conditions go over to
  \begin{align}
  &  \widetilde{\alpha}(0) =
  f^{*}\left(\widetilde{\alpha}(1)\rho(k)\right) \equiv
  f^{*}\, . \, \widetilde{\alpha}(1)\rho(k) \nonumber\\
&  \widetilde{\beta}(0) =  
f^{*}\left(\rho(k)^{-1}\widetilde{\beta}(1)\right) \equiv
f^{*}\, . \, \rho(k)^{-1}\widetilde{\beta}(1)
  \end{align}
where here now $k \equiv k(t=1)$.
The path integral can now be evaluated by discretising the interval as we did for the
$\eta$-$\overline{\eta}$ system in Section \ref{MRSTS1}. In this way
we arrive at
\be
\int D\widetilde{\beta}(0) \delta{ \left( (1- f^{*}\, .\, 
    \rho(k)^{-1})\widetilde{\beta}(0)  \right)} =
\Det_{ \Omega^{i}(N,\mathbb{E})}{(1- f^{*}\, . 
    \rho(k)^{-1})}^{\pm 1}
\ee
the power depending on the statistics of $\alpha$ and $\beta$. This
only differs from the situation on $N \times S^{1}$ by the pull back
with respect to $f$ which acts both on the holonomies as well as the
space of forms $\Omega^{i}(N, \mathbb{E})$. Consequently, we may make use of the
work done in Section \ref{PI-Massive-RST} to arrive at
\be
Z_{N_f} =  \prod_{i=0}^{n-1} Z_{i} 
= \prod_{i=0}^{n-1}\Det_{\Omega^{i}(N, \mathbb{E})}{(1- f^{*}\, . \, 
    \rho(k)^{-1})}^{(-1)^{p+1+i}} \label{Z}\;\;.
\ee
The appropriate regularised form of the determinant can be obtained 
by writing (as usual) $\log\Det = \Tr \log$, 
the heat kernel regularised logarithm being
\be
\lim_{\eps \rightarrow 0}
\sum_{i=0}^{n-1}(-1)^{i}\Tr_{\Omega^{i}(N,\mathbb{E})  }{\left(
    \ln{\left(1-f^{*}\, . \, \rho(k)^{-1}\right)}\, \ex{-\eps \Delta_{A}}
\right)} \label{local-form-Lefschetz} 
\ee 
Compared with the analogous calculation on $N \times S^1$ performed in 
detail in the previous section, here we now also have a
`twist' coming from taking into account the action of the finite order
diffeomorphism on the spaces of forms. It is also convenient to use an
$f$-invariant metric (since $f$ is of finite order, this can always be 
achieved for any seed metric $g_N$ by `averaging') and we do so.

When $\phi$ and $\mu$ are constant on $N$, we may separate the action
on the space of forms as
\be
\lim_{\eps \rightarrow 0}
\sum_{i=0}^{n-1}(-1)^{i}\Tr_{\Omega^{i}(N,\mathbb{E})}{\left(
    \ln{(1-\rho(k)^{-1}\, f^{*})}\, \ex{-\eps \Delta_{A}}
\right)}
\ee 
Note that since $(f^*)^r$ is an isometry, one has  
\be
\lim_{\eps \rightarrow 0} \sum_{i=0}^{n-1}(-1)^{i}\Tr_{\Omega^{i}(N,\mathbb{E})}{\left(
      \ex{-\eps \Delta_{A}} \, (f^*)^r \right)} =
  \sum_{i=0}^{n-1}(-1)^{i}\Tr_{\mathrm{H}^{i}_{A}(N,\mathbb{E})}{\left( (f^*)^r\right)}
\ee
Thus, on expanding the logarithm we understand that we may replace the
signed sum over the spaces of forms with the signed sum of the
cohomology groups so that on putting these pieces together we find
\be
\tau_{M}(\phi+ \mu, \mathbb{E}) =
\prod_{i=0}^{n-1}\Det_{\mathrm{H}^{i}(N,\EE)}{(1- 
    \rho(k)^{-1} \otimes f^{*}_{i})}^{(-1)^{i}}
  \ee
This is already a generalisation of Fried's formula (\ref{fried-map}) as the
  connection does not have to be the pullback of a connection on the
  base (but could include an arbitrary flat component $A_N$ along $N$)
  and includes the $\mathbb{R}_{+}$ connection (or a mass,  as one prefers).

In general, when $\phi + \mu$ is not constant on $N$ evaluating
(\ref{Z}) may be laborious. One way is to split the $\Omega^{i}(N,\EE)$ in
terms of eigenspaces of $f$ (with eigenvalues characters of the finite cyclic 
group generated by $f$) as
\be
\Omega^{i}(N, \mathbb{E}) =\bigoplus_{s=0}^{n-1}\, \Omega^{i}_{(s)}(N, \mathbb{E}), \quad
  f^{*}\omega^{i}_{(s)} = \chi_{s}(f) \omega^{i}_{(s)} \;\; \mathrm{for}
  \;\; \omega^{i}_{(s)} \in \Omega^{i}_{(s)}(N, \mathbb{E})
  \ee
  Also with the notation at hand we must understand that there is a corresponding decomposition of $\rho$, with $\rho_{(s)}$ acting as 
  \be
  \rho_{(s)}: \Omega^{i}_{(r)}(N, \mathbb{E}) \rightarrow
  \Omega^{i}_{(r+s)}(N, \mathbb{E} )
  \ee
Then the determinants in (\ref{Z}) may be expressed in terms of
products of determinants over the spaces $
\Omega^{i}_{(s)}(N, \mathbb{E})$. One would then find a result of the 
general (but enlightening) form 
\be
Z = \prod_{s=0}^{n_f -1}\left(\prod_{i=0}^{n-1}\Det_{
    \Omega^{i}_{(s)} (N, \mathbb{E})}{\left(T_{s}(f, \rho(k))
      \right)}^{(-1)^{p+1+i}} \right) \label{Z2}
\ee
At this point a McKean-Singer type argument could be used for the
alternating signed sum over the $\Omega^{\bullet}_{(s)}$ for each
eigenspace separately.

  \subsection*{Acknowledgements}
 M. Kakona thanks the External Activities Unit of the ICTP for a
 Ph.D. grant at EAIFR in the University of Rwanda and the HECAP group at the ICTP for
 supporting a visit to complete this work. The group of M. Blau 
is supported by the NCCR SwissMAP (The Mathematics of Physics) of the Swiss Science Foundation. 

\appendix

\section{Properties of the Massive Ray-Singer Torsion}\label{mRST-props}

Here we 
establish two properties of the massive
Ray-Singer Torsion mentioned in Section \ref{submrst}, namely
\begin{enumerate}
\item the triviality of the massive Ray-Singer Torsion on even-dimensional 
manifolds $M$ \eqref{even-RST}
\be
\dim M = 2\ell \quad\Ra\quad 
      \tau_{M}(A, \mathbb{E},g_M, m) = 1 \label{even-RST2}
\ee
\item a product formula for the massive Ray-Singer Torsion on product manifolds
$M\times N$, 
when the flat connection $A$ is of the 
form $A \in \Omega^{1}(M, \lg) \otimes \Omega^{0}(N)$, 
namely \eqref{product-RST}
      \be
      \tau_{M\times N}(A, \mathbb{E}, m) = \tau_{M}(A, \mathbb{E},
      m)^{e(N)} \, . \, \tau_{N}(m)^{e(M).\rk{\mathbb{E}}} 
\label{product-RST3}
      \ee
\end{enumerate}
Both of these properties are strict analogues of those for the
traditional Ray-Singer Torsion and, indeed, the proofs of these
properties can also be modelled on the traditional proofs.

\subsection{Triviality of the Massive Ray-Singer Torsion in Even Dimensions}

We already outlined one proof of this assertion (based on the
generalisation of the original argument of Ray and Singer to the
massive case) in Section \ref{submrst}. Here we will generalise an
alternative argument to the massive case, namely one originally
given in \cite{BT-BF} and inspired by the path integral representation
of the Ray-Singer Torsion.

First of all, we recall the definition of the massive Ray-Singer Torsion, 
namely 
\be
\tau_{M}(A, \mathbb{E}, g_M,m)= |m|^{e(M).\rk(\mathbb{E}). \dim{M}/2} \, 
\prod_{i=0}^{\dim M} \left(\Det_{\Omega^{i}(M, \mathbb{E})}{(\Delta_{A}+ m^{2}
    )}\right)^{(-1)^{i+1}i/2} \label{mass-RST2}
\ee
Using Hodge duality, in the even-dimensional case $\dim M = 2\ell$ this 
can be rewritten as (suppressing the arguments of $\tau_M$, and abbreviating
$\Omega^i(M,\EE) \equiv \Omega^i$)
\be
\tau_M = |m|^{e(M)\rk(\EE)\ell}
\left(\Det_{\Omega^{\ell}}(\Delta_{A}+ m^{2})\right)^{(-1)^{\ell+1}\ell/2}
\prod_{i=0}^{\ell-1} \left(\Det_{\Omega^{i}}{(\Delta_{A}+ m^{2}
    )}\right)^{(-1)^{i+1}\ell} 
\label{massrsteven}
\ee
On the other hand, in \cite{BT-BF} 
a simple scaling argument based on the path integral 
was used to establish the identity (Proposition 4 in \cite{BT-BF}, in present
notation)
\be
\Det'_{\Omega^{\ell}}f(\Delta_{A}) = 
\prod_{i=0}^{\ell-1} \left(\Det'_{\Omega^{\ell-i-1}}{f(\Delta_{A})}
\right)^{2(-1)^{i}} 
\ee
where $f(\Delta_A)$ is some function of the Laplacian, and the notation
$\Det'$ indicates that the harmonic modes
of $\Delta_A$ were projected out. This formula can be applied to the massive Laplacian, i.e.\ to the simple function 
\be
f(\Delta_{A}) = \Delta_{A} + m^{2} \;\;,
\ee
but in order to conform with our definition of the determinant of 
$\Delta_A+m^2$, 
which also included the harmonic modes of $\Delta_A$, each determinant 
appearing in the above identity needs to be divided explicitly by the
contribution of the harmonic modes to the determinant, i.e.\ 
\be
\Det'_{\Omega^{i}}(\Delta_{A}+m^2)  = 
m^{-2\dim H^i_A(M,\EE)} 
\Det_{\Omega^{i}}(\Delta_{A}+m^2) \;\;.
\ee
Collecting all the powers of the mass $m$, one then arrives at the identity 
\be
\Det_{\Omega^{\ell}}(\Delta_{A}+m^2) = m^{2(-1)^\ell e(M)\rk(E)}
\prod_{i=0}^{\ell-1} \left(\Det_{\Omega^{\ell-i-1}}{(\Delta_{A}+ m^2)}
\right)^{2(-1)^{i}} 
\ee
This is now the `massive' generalisation of Proposition 4 of \cite{BT-BF}. 
Comparing this with the expression 
\eqref{massrsteven} for the massive Ray-Singer Torsion obtained above, we
conclude that $\tau_M=1$.

\subsection{Product Formula for the Massive Ray-Singer Torsion on $M\times N$}

We now turn to a proof of the product formula \eqref{product-RST3} for the 
massive Ray-Singer Torsion. In \cite{BT-CS} we gave a slight
generalisation of Theorem 2.5 in \cite{Ray-Singer} which we extend
here to the massive Ray-Singer Torsion 
$\tau_{M}(A, \mathbb{E}, m)$. Suppose $M$ and $N$ are compact, closed and
oriented manifolds and equip $M \times N$ with the product
metric. Let $E$ be a vector bundle over $M \times N$ which
allows for connections with local form $A \in \Omega^{1}(M, \lg)
\otimes \Omega^{0}(N)$, up to gauge equivalence. (Thus in contrast
to Ray and Singer, we do not require $\pi_1(N)$ to be trivial; and 
we also do not need $\EE$ and $A$ to be pullbacks from $M$ to $M\times N$.)
Then one has 
      \be
      \tau_{M\times N}(A, \mathbb{E}, m) = \tau_{M}(A, \mathbb{E},
      m)^{e(N)} \, . \, \tau_{N}(m)^{e(M).\rk{\mathbb{E}}} 
\label{RSTprod}
      \ee
Note that here $\tau_N(m)$ simply refers to the massive Ray-Singer Torsion
of the untwisted Laplacian $\Delta + m^{2}$ on differential forms on $N$. 

By the previous result, the only interesting case is when 
$\dim M$ is even and $\dim N$ is odd (or perhaps vice-versa), 
but we will give the proof in general. 
The proof is essentially identical to that of Ray and 
Singer, the difference being that we allow for zero eigenvalues of the
Laplacian. With the product metric one has that the twisted Laplacian 
acts `diagonally', 
\[
\Delta_{A} \left( \omega_{M} \otimes \omega_{N} \right) =
(\Delta_{A}\omega_{M})\otimes \omega_{N}  +\omega_{M}\otimes ( \Delta
\omega_{N}), \;\;  \omega_{M} \in \Omega(M, \mathbb{E}), \; \omega_{N} \in \Omega(N)
\]
The $\zeta$ function for the massive Laplacian acting on
$\Omega^{r}(M\times N,\EE)$ is thus
\[
\zeta_{r}(M\times N,\EE)(s) = \sum_{\lambda, \nu} \sum_{p+q=r} (\lambda +
\nu + m^{2})^{-s} M_{p}(\lambda, M)\, M_{q}(\nu, N)
\]
with $M_{p}(\lambda, M)$ and $M_{q}(\nu, N)$ being the multiplicity of the eigenvalues
$\lambda$ and $\nu$ of the Laplacian on the spaces of forms
$\Omega^{p}(M,\EE)$ and $\Omega^{q}(N)$ respectively. The alternating
sum which goes into the definition of the torsion is
\be
\begin{aligned}
\sum_{r=0}^{\dim(M\times N)}& (-1)^{r}r\, \zeta_{r}(M\times N,\EE)(s) \\
&=  \sum_{\lambda, \nu}(\lambda +
\nu + m^{2})^{-s} \sum_{p=0}^{\dim M} \sum_{q=0}^{\dim N} (-1)^{p+q} (p+q)
M_{p}(\lambda, M) M_{q}(\nu, N) \\
 & = \sum_{\lambda, \nu}(\lambda +
\nu + m^{2})^{-s}  \sum_{p=0}^{\dim M} (-1)^{p} p M_{p}(\lambda,
M)\sum_{q=0}^{\dim N} (-1)^{q} M_{q}(\nu, N) \\
&+ \sum_{\lambda, \nu}(\lambda + \nu + m^{2})^{-s} 
\sum_{p=0}^{\dim M} (-1)^{p}  M_{p}(\lambda,
M)\sum_{q=0}^{\dim N} (-1)^{q} q M_{q}(\nu, N) 
\end{aligned}
\ee

Ray and Singer
show that (we have allowed $\lambda=0$)
\[
\sum_{p=0}^{\dim M} (-1)^{p} M_{p}(\lambda, M) = e(M).\rk(\mathbb{E}) \delta_{\lambda 0}\;\;,
\;\;\; \sum_{q=0}^{\dim N} (-1)^{q} 
M_{q}(\nu, N) = e(N)\delta_{\nu 0} 
\]
(non-zero
eigenvalues of the Laplacian do not contribute to the Euler characteristic and
so their total signed sum must be zero) whence
\be
\begin{aligned}
\sum_{r=0}^{\dim(M\times N)}& (-1)^{r}r\, \zeta_{r}(M\times N,\EE)(s)\\
&= e(N)\sum_{\lambda}(\lambda + m^{2})^{-s} 
\sum_{p=0}^{\dim M} ( -1)^{p} p M_{p}(\lambda, M) \\
&+ e(M).\rk(\mathbb{E}) \sum_{\nu}(\nu + m^{2})^{-s}
\sum_{q=0}^{\dim N} (-1)^{q} q M_{q}(\nu, N) \\
&=  e(N) \sum_{r=0}^{\dim M} (-1)^{r}r\, \zeta_{r}(M,\EE)(s) +
e(M).\rk(\mathbb{E}) \sum_{r=0}^{\dim N} (-1)^{r}r\,  \zeta_{r}(N)(s) 
\end{aligned}
\ee
Taking the derivative at $s=0$ now yields the desired product formula 
\eqref{RSTprod}.

\rnc{\Large}{\normalsize}

\end{document}